\documentclass[12pt]{article}
\pdfoutput=1
\usepackage[english]{babel}
\usepackage{amssymb,slashed,latexsym,amsmath,multirow,color}
\usepackage{slashed}
\usepackage{tikz}
\usetikzlibrary{shapes,shadows,arrows}
\usepackage[font=footnotesize,labelsep=newline,labelfont=sc,justification=centering,position=top]{caption}

\usepackage{tikz}
\usetikzlibrary{backgrounds}
\usetikzlibrary{shapes,shadows,arrows}
\tikzstyle{every picture}=[level distance = 8mm, baseline=-0.5ex]
\tikzstyle{prop}=[shape=circle,minimum size=6mm, draw=black!80, fill=green!30]

\pdfoutput=1
\usepackage[font=footnotesize,labelsep=newline,labelfont=sc,justification=centering,position=top]{caption}

\numberwithin{equation}{section}

\newcommand{\opsi}{\bar{\psi}}

\usepackage{hyperref}

\newcommand{\ft}[2]{{\textstyle\frac{#1}{#2}}}
\def\rmi{{\rm i}}
\newsavebox{\uuunit}
\sbox{\uuunit}
    {\setlength{\unitlength}{0.825em}
     \begin{picture}(0.6,0.7)
        \thinlines
        \put(0,0){\line(1,0){0.5}}
        \put(0.15,0){\line(0,1){0.7}}
        \put(0.35,0){\line(0,1){0.8}}
       \multiput(0.3,0.8)(-0.04,-0.02){12}{\rule{0.5pt}{0.5pt}}
     \end {picture}}

\newcommand{\SU}{\mathop{\rm SU}}

\def\be{\begin{equation}}
\def\ee{\end{equation}}
\def\ba{\begin{array}}
\def\ea{\end{array}}
\def\bea{\begin{eqnarray}}
\def\eea{\end{eqnarray}}
\def\bd{\begin{displaymath}}
\def\ed{\end{displaymath}}

\def\nn{\nonumber}

\def\g{\gamma}

\def\d{\delta}

\def\e{\epsilon}

\def\f{\phi}
\def\vf{\varphi}

\def\p{\psi}

\def\l{\lambda}
\def\L{\Lambda}
\def\m{\mu}
\def\n{\nu}
\def\r{\rho}
\def\s{\sigma}

\def\t{\tau}

\def\o{\omega}

\def\nn{\nonumber}

\begin{document}

\begin{flushright}
\hfill{ \
\ \ \ \ MIFPA-12-41\ \ \ \ }
\end{flushright}
\vskip 1.2cm
\begin{center}
{\Large \bf An Off-Shell Formulation for Internally Gauged \texorpdfstring{$D=5,\,{\cal{N}}=2$}{} Supergravity from Superconformal Methods
 }
\\

\end{center}
\vspace{25pt}
\begin{center}
{\Large {\bf }}

 \vspace{15pt}

\textbf{F. Coomans$^1$, M. Ozkan$^2$}

\vspace{20pt}

\vskip .2truecm  \centerline{{\small $^1$Instituut voor Theoretische Fysica, Katholieke Universiteit
 Leuven,}}
      \centerline{{\small  Celestijnenlaan 200D B-3001 Leuven, Belgium}}
\vskip .2truecm
{email: {\tt frederik.coomans@fys.kuleuven.be}}
\vskip .2truecm
\centerline{{\small  $^2$ George and Cynthia Woods Mitchell
Institute for Fundamental Physics and Astronomy,}} \centerline{\small Texas
A\&M University, College Station, TX 77843, USA}  \vspace{6pt}

{email: {\tt mozkan@tamu.edu}}
\vspace{40pt}

\underline{ABSTRACT}
\end{center}

We use the superconformal method to construct a new formulation for pure off-shell $D=5$, ${\cal{N}}=2$ Poincar\'e supergravity and present its internal gauging. The main difference between the traditional formulation and our new formulation is the choice of the Dilaton Weyl Multiplet as the background Weyl Multiplet and the choice of a Linear compensating Multiplet. We do not introduce an external Vector Multiplet to gauge the theory, but instead use the internal vector of the Dilaton Weyl Multiplet. We show that the corresponding on-shell theory is Einstein-Maxwell supergravity. We believe that this gauging method can be applied in more complicated scenarios such as the inclusion of off-shell higher derivative invariants.

\vspace{15pt}

\thispagestyle{empty}

\vspace{15pt}

\thispagestyle{empty}

\newpage

\tableofcontents


\newpage


\section{Introduction}

Pure on-shell five dimensional supergravity with eight supercharges was first introduced in \cite{Cremmer:1980gs}, its gauging was investigated in detail in \cite{Gunaydin:1984ak, Gunaydin:1999zx} and general matter couplings and their geometrical aspects were studied in \cite{Ceresole:2000jd, Bergshoeff:2004kh, Bergshoeff:2002qk}. To study general matter couplings it is useful to work in an off-shell superconformal setting and, eventually, gauge fix the redundant conformal symmetries. To this end, the ${\cal{N}}=2$ superconformal program was initiated in \cite{Bergshoeff:2001hc, Fujita:2001kv}.

The pure on-shell theory consists of the metric, graviphoton and gravitino. We denote an off-shell pure theory as a theory with minimal field content on which the super Poincar\'e algebra closes off-shell and which reduces, upon going on-shell and decoupling the matter fields, to the pure on-shell theory. The off-shell nature of the theory implies that, in addition to the metric, graviphoton and gravitino, they also contain auxiliary fields. Off-shell formulations are constructed most easily by using the method of superconformal tensor calculus. Different off-shell formulations can correspond to one physical on-shell theory. They can differ in the compensator multiplet used to compensate for the redundant conformal symmetries or in the choice of Weyl multiplet. In 5D superconformal tensor calculus there are two possible choices for the Weyl multiplet: the Standard Weyl and the Dilaton Weyl Multiplet. These multiplets contain the same superconformal gauge fields but differ in their matter sector: a scalar $D$, an antisymmetric tensor $T_{ab}$ and a symplectic Majorana spinor $\chi^i$ for the Standard Weyl Multiplet and a scalar $\sigma$, a vector $C_\m$, a 2-form $B_{\m\n}$ and a symplectic Majorana spinor $\psi^i$ for the Dilaton Weyl Multiplet. In \cite{Hanaki:2006pj} a pure off-shell 5D theory with eight supercharges is written down using the Standard Weyl Multiplet and a Hypermultiplet compensator, and in \cite{Zucker:1999ej}, a Nonlinear Multiplet was used as compensator instead of a Hypermultiplet. In this paper we will write down a different off-shell pure theory using the Dilaton Weyl Multiplet and a Linear Multiplet compensator.

Off-shell formulations are of great use when studying higher order curvature extensions of the theory. Off-shell formulations for curvature squared invariants have been constructed in 5 dimensions in \cite{Hanaki:2006pj, Bergshoeff:2011xn}. Unlike an effective supergravity Lagrangian of a compactified string theory which has higher-order correction terms in $\alpha'$ and which is supersymmetric only order by order in $\alpha'$, these invariants can be added to a pure off-shell supergravity theory, and they are exactly supersymmetric. The invariant constructed in \cite{Hanaki:2006pj} is the supersymmetric completion of the Weyl tensor squared and makes use of an external Vector Multiplet manifested in the appearance of a mixed gauge-gravitational Chern-Simons (CS) term $A\wedge \text{tr}(R \wedge R)$, where $A$ denotes the external vector. It is constructed in the background of the Standard Weyl Multiplet. The invariant constructed in \cite{Bergshoeff:2011xn} is the supersymmetric completion of the Riemann tensor squared. It is constructed in the background of the Dilaton Weyl Multiplet and is based on a map between the Yang-Mills Multiplet and the Dilaton Weyl Multiplet. It does not require an external Vector Multiplet, but rather it has a purely gravitational CS term $C\wedge \text{tr}(R \wedge R$), where $C$ denotes the internal vector of the Dilaton Weyl Multiplet.

In \cite{Bergshoeff:2001hc} it was shown that the Dilaton Weyl Multiplet can be obtained by solving the equations of motion for a Vector Multiplet coupled to the Standard Weyl Multiplet. In this paper our purpose is to exploit this connection between the Weyl multiplets to obtain an off-shell Lagrangian for 5D minimal supergravity in the background of a Dilaton Weyl Multiplet in which the $U(1)$ R-symmetry is gauged dynamically by the internal vector $C$, i.e. the vector field $C$ has a kinetic term in the Lagrangian. This forms a basis for a future study of the Weyl squared invariant in a Dilaton Weyl background with the external vector $A$ replaced by the internal one $C$. A similar discussion in 4D can be found in \cite{deWit:2006gn}.

The paper is built up as follows. In section \ref{section: multiplets} we summarize the 5D superconformal calculus and introduce the Standard Weyl Multiplet as well as two types of matter multiplets: the Linear Multiplet and the Vector Multiplet. In section \ref{section: actions} we construct invariant actions for the Linear and Vector Multiplet. In section \ref{ungaugedtheory} we construct 5D pure off-shell supergravity by coupling the Standard Weyl Multiplet to a Linear compensator. This procedure is very similar to the one used in 6D in \cite{Coomans:2011ih}. We then add a superconformal abelian Vector Multiplet action to the Linear Multiplet action and compute the field equations for the Vector Multiplet components. These equations allow us to solve for the matter fields of the Standard Weyl Multiplet ($D$, $T_{ab}$ $\chi^i$) in terms of the fields of the Vector Multiplet\footnote{We suggestively denote the fields of the Vector Multiplet with $\sigma$, $C_\m$ and $\psi^i$. The Vector Multiplet also has an auxiliary $Y^{ij}$, but we solve for this auxiliary and use its value in the expressions for $D$, $T_{ab}$ and $\chi^i$.} ($\sigma$, $C_\m$, $\psi^i$) plus an additional 2-form ($B_{\m\n}$). Using these expressions in the Lagrangian we obtain the action for the Linear Multiplet in the background of the Dilaton Weyl Multiplet. Gauge fixing the redundant conformal symmetries leads to off-shell Poincar\'e supergravity which has apart from the graviton and the gravitino, a vector field, a 2-form gauge field, a dilaton, a symplectic Majorana spinor and a number of auxiliary fields. 

In section \ref{gaugedtheory} we develop an off-shell method to gauge the theory. We start with an off-shell action consisting of the Linear Multiplet action, the Vector Multiplet action and a coupling between the Vector and Linear Multiplet, all in the background of the Standard Weyl Multiplet. Then we compute, as in the ungauged case, the field equations for the Vector Multiplet components to obtain expressions for the Standard Weyl matter fields. We notice that these expressions get deformed by the Vector-Linear coupling. After using these expressions in the Lagrangian and gauge fixing the conformal symmetries we obtain an off-shell expression for $U(1)$-gauged supergravity. After eliminating the auxiliary fields and dualizing the 2-form $B_{\m\n}$ to a vector $\tilde{C}_{\m}$, we show that the resulting theory is Einstein-Maxwell supergravity gauged by a linear combination of $C_{\m}$ and $\tilde{C}_{\m}$. This theory agrees completely with the one constructed in \cite{Gunaydin:1984ak} via the Noether procedure. Finally, we show that we can consistently eliminate $\sigma$, $\psi$ and $\tilde{C}_{\m}$. The resulting on-shell theory is minimal gauged 5D supergravity \cite{Gunaydin:1984ak} consisting of the graviton, the graviphoton and the gravitino. 


\section{Superconformal Multiplets}\label{section: multiplets}
In this section, we will recall the elements of ${\cal{N}} = 2, D=5$ superconformal tensor calculus constructed in \cite{Bergshoeff:2001hc, Fujita:2001kv}. In the first subsection \ref{ss: standardweyl} we introduce the gauge multiplet of the ${\cal{N}} = 2, D=5$ superconformal group: the Standard Weyl Multiplet. In the last two subsections, \ref{ss: vector} and \ref{ss: linear}, we introduce two types of matter multiplets, the Vector Multiplet and the Linear Multiplet.

\subsection{The Standard Weyl Multiplet} \label{ss: standardweyl}
The ${\cal{N}} = 2, D=5$ superconformal tensor calculus is based on the superconformal algebra\footnote{The notation $F^p(4)$ refers to a compact form of $F(4)$ with bosonic subalgebra $SO(7-p,p)$.} $F^{2}(4)$ with the generators
\begin{eqnarray}
P_{a},\ \ \ M_{ab},\ \ \  D,\ \ \  K_{a},\ \ \  U_{ij},\ \ \  Q_{\alpha i},\ \ \  S_{\alpha i}\,, \label{gen}
\end{eqnarray}
where $a,b,\ldots$ are Lorentz indices\footnote{We use the conventions of \cite{Bergshoeff:2001hc, Freedman:2012zz}. In particular, the spacetime signature is $(-,+,+,+,+)$ and $\bar{\psi}^i\gamma_{(n)}\chi^j=t_n\bar{\chi}^j\gamma_{(n)}\psi^i$ with $t_0=t_1=-t_2=-t_3=1$. When $SU(2)$ indices on spinors are omitted, northwest-southeast contraction is understood.}, $\alpha$ is a spinor index and $i = 1,2$ is an $SU(2)$ index. Here $M_{ab}$ and $P_{a}$ are the usual Poincar\'e generators, $D$ is the generator for dilatations, $K_{a}$ generates special conformal boosts, $U_{ij}$ is the $SU(2)$ generator and $Q_{\alpha i}$ and $S_{\alpha i}$ are the supersymmetry and conformal supersymmetry generators respectively. \\
For each of the generators above we now introduce the following gauge fields
\begin{eqnarray}
h_{\mu}{}^A\equiv\{e_{\m}{}^{a},\ \ \ \omega_{\m}{}^{ab}, \ \ \ b_{\m}, \ \ \ f_{\m}{}^{a}, \ \ \ V_{\m}^{ij}, \ \ \ \psi_{\m}^{i}, \ \ \ \phi_{\m}^{i}\}\,,
\label{gf} 
\end{eqnarray}
where $\m, \n, \ldots$ are world vector indices. Using the structure constants $f_{AB}{}^C$ of the superconformal algebra (given e.g. in appendix B of \cite{Bergshoeff:2001hc}) and the basic rules
\bea
\delta h_\m^A&=&\partial_\m \epsilon^A+\epsilon^Ch_\m^B f_{BC}{}^A\,, \nn \\
R_{\m\n}{}^A&=&2\partial_{[\m}h_{\n]}^A+h_\n^Ch_\m^Bf_{BC}{}^A\,,
\eea
one can easily write down the linear transformation rules and the linear curvatures $R_{\m\n}{}^A$ of the superconformal gauge fields given in (\ref{gf}). The linear transformations given in \cite{Bergshoeff:2001hc} satisfy the $F^2(4)$ superalgebra, thus resulting in a gauge theory of $F^2(4)$ since we have not related the generators $P_{a}, M_{ab}$ to the diffeomorphisms of spacetime. This problem can be solved by imposing the so-called curvature constraints \cite{Bergshoeff:2001hc}. These constraints determine the gauge fields $\omega_{\m}{}^{ab}$, $\phi_\m^i$ and $f_\m{}^a$ in terms of the independent gauge fields $e_\m{}^a$, $\psi_\m^i$, $b_\m$, $V_\m^{ij}$ and, in addition, achieve maximal irreducibility of the superconformal gauge field configuration.

A simple counting argument shows that the superconformal gauge fields, after imposing the conventional constraints, represent 21 + 24 off-shell degrees of freedom and therefore cannot represent a supersymmetric theory. Additional matter fields $T_{\m\n} (10), D (1)$ and $\chi^{i} (8)$ must be added to the gauge fields in order to obtain an off-shell closed multiplet \cite{Bergshoeff:2001hc, Fujita:2001kv}. This multiplet is known as the Standard Weyl Multiplet.

Starting from the linear transformation rules of the superconformal fields, the linear curvatures $R_{\m\n}{}^A$ and the matter fields $T_{\m\n}$, $D$, and $\chi^{i}$ we can construct the full nonlinear ${\cal{N}} = 2, D=5$ Weyl Multiplet by applying an iterative procedure (described in detail for 6 dimensions in \cite{Bergshoeff:1985mz}). The results are \cite{Bergshoeff:2001hc} (we only give $Q$, $S$ and $K$ transformations):
\bea
\d e_\m{}^a   &=&  \ft 12\bar\e \g^a \psi_\m  \nn\, ,\\
\d \psi_\m^i   &=& (\partial_\m+\tfrac{1}{2}b_\m+\tfrac{1}{4}\omega_\m{}^{ab}\g_{ab})\e^i-V_\m^{ij}\e_j + \rmi \g\cdot T \g_\m
\e^i - \rmi \g_\m
\eta^i  \nn\, ,\\
\d V_\m{}^{ij} &=&  -\ft32\rmi \bar\e^{(i} \phi_\m^{j)} +4
\bar\e^{(i}\g_\m \chi^{j)}
  + \rmi \bar\e^{(i} \g\cdot T \psi_\m^{j)} + \ft32\rmi
\bar\eta^{(i}\psi_\m^{j)} \nn\, ,\\
\d T_{ab} &=& \tfrac12 \rmi\bar\e \g_{ab} \chi - \tfrac3{32} \rmi \bar\e \widehat{R}_{ab}(Q)\,, \nn
\eea
\bea
\d \chi^i &=& \tfrac14 \e^i D - \tfrac1{64} \g \cdot  \widehat{R}^{ij}(V) \e_j + \tfrac18 \rmi \g^{ab}\slashed{\mathcal{D}}T_{ab}\e^i - \tfrac18 \rmi \g^a  \mathcal{D}^b T_{ab} \e^i \nn\\
&& - \tfrac14 \g^{abcd}T_{ab} T_{cd} \e^i + \tfrac16 T^2 \e^i + \tfrac14 \g \cdot T \eta^i\,, \nn\\
\d D &=& \bar\e \slashed{\mathcal{D}}\chi - \tfrac53 \rmi \bar\e \g \cdot T \chi - \rmi \bar\eta \chi\,, \nn\\
 \d b_\m       &=& \ft12 \rmi \bar\e\phi_\m -2 \bar\e\g_\mu \chi +
\ft12\rmi \bar\eta\psi_\mu+2\Lambda _{K\mu } \,,
\label{dwmtr}
\eea
where
\bea 
\mathcal{D}_\m\chi^i&=&(\partial_\mu - \tfrac72 b_\mu +\tfrac14 \omega_\mu{}^{ab} \g_{ab})\chi^i -V_\mu^{ij}\chi_j
 - \tfrac14 \p_\m^i D  + \tfrac1{64} \g \cdot  \widehat{R}^{ij}(V) \p_{\m j}\nn\\
&& - \tfrac18 \rmi \g^{ab}\slashed{\mathcal{D}}T_{ab}\p_\m^i + \tfrac18 \rmi \g^a  \mathcal{D}^b T_{ab} \p_\m^i  + \tfrac14 \g^{abcd}T_{ab} T_{cd} \p_\m^i - \tfrac16 T^2 \p_\m^i - \tfrac14 \g \cdot T \phi_\m^i\,, \nn \\
\mathcal{D}_\m T_{ab}&=&\partial_\m T_{ab}-b_\m T_{ab}-2\omega_\mu{}^c{}_{[a}T_{b]c}-\tfrac{1}{2}i\bar{\p}_\m\g_{ab}\chi+\tfrac{3}{32}i\bar{\p}_\m\widehat{R}_{ab}(Q)\,.
\eea
The relevant modified curvatures are
\bea
\widehat{R}_{\m\n}{}^{ab}(M)&=&2\partial_{[\m}\o_{\n ]}{}^{ab}+2\o_{[\m}{}^{ac}\o_{\n ]c}{}^{b} + 8 f_{[\m}{}^{[a}e_{\n ]}{}^{b]}+\rmi \bar\p_{[\m}\g^{ab}\p_{\n ]} + \rmi \bar\p_{[\m}\g^{[a} \g \cdot T \g^{b]}\p_{\n ]}  \nn\\
&& +\bar\p_{[\m} \g^{[a} \widehat{R}_{\n ]}{}^{b]}(Q)+\tfrac12 \bar\p_{[\m}\g_{\n ]} \widehat{R}^{ab}(Q) -8 \bar\p_{[\m} e_{\n ]}{}^{[a} \g^{b]}\chi+i\bar{\phi}_{[\m} \g^{ab} \p_{\n]}\nn\,, \\
\widehat{R}_{\m\n}{}^{ij}(V)&=&2\partial_{[\m} V_{\n]}{}^{ij} -2V_{[\m}{}^{k( i}
V_{\n ]\,k}{}^{j)} {-3\rmi}{\bar\f}^{( i}_{[\m}\p^{j)}_{\n ]}  - 8 \bar{\p}^{(i}_{[\m}
\g_{\n]} \chi^{j)} - \rmi \bar{\p}^{(i}_{[\m} \g\cdot T \psi_{\n]}^{j)}   \,, \nn\\
\widehat{R}_{\m\n}^i(Q)&=&2\partial_{[\m}\p_{\n]}^i+\frac{1}{2}\o_{[\m}{}^{ab}\g_{ab}\p_{\n]}^i+b_{[\m}\p_{\n]}^i-2V_{[\m}^{ij}\p_{\n] j}-2i\g_{[\m}\phi_{\n]}^i+2i\g\cdot T\g_{[\m}\p_{\n]}^i\,.
\eea 
As mentioned before, the dependent fields, which relate the generators $P_{a}, M_{ab}$ to the diffeomorphisms of spacetime, are completely determined by the following curvature constraints
\bea
\label{constraints}
R_{\mu\nu}{}^a(P) = 0\,, \nn\\
e^{\nu}{}_b \widehat R_{\mu\nu}{}^{ab} (M) = 0\,, \nn\\
\g^\m \widehat R_{\m\n}{}^i (Q) = 0\,.
\label{cc}
 \eea
Notice that our choices for the above constraints are not unique, i.e. one can impose different constraints by adding further terms to (\ref{cc}). However such additional terms only amount to redefinitions of the dependent fields defined below. Using the curvature constraints we identify $\omega_{\m}{}^{ab}$, $\phi_\m^i$ and $f_\m{}^a$ in terms of the other gauge fields and matter fields
 \bea \o_\m{}^{ab}
&=& 2 e^{\n[a} \partial_{[\m} e_{\n]}^{~b]} - e^{\n[a} e^{b]\s} e_{\m c}
\partial_\n e^{~c}_\s
 + 2 e_\m^{~~[a} b^{b]} - \ft12 \bar{\p}^{[b} \g^{a]} \p_\m - \ft14
\bar{\p}^b \g_\m \p^a \,,\nonumber\\
\f^i_\m &=& \ft13\rmi \g^a \widehat{R}^\prime _{\m a}{}^i(Q) - \ft1{24}\rmi
\g_\m \g^{ab} \widehat{R}^\prime _{ab}{}^i(Q)\,, \label{transfDepF} \\
f^a_\m &=&  - \ft16{\cal R}_\mu {}^a + \ft1{48}e_\mu {}^a {\cal R}\,, \nonumber
\label{cf1}
\eea
where ${\cal R}_{\mu \nu }\equiv \widehat{R}_{\mu \rho }^{\prime~~ab}(M) e_b{}^\rho
e_{\nu a}$ and ${\cal R}\equiv {\cal R}_\mu {}^\mu$. The notation $\widehat{R}'(M)$ and $\widehat{R}'(Q)$ indicates that we have omitted the $f_\m{}^a$ dependent term in $\widehat{R}(M)$ and the $\phi_\m^i$ dependent term in $\widehat{R}(Q)$. The constraints imply through Bianchi identities further relations
between the curvatures. The Bianchi identities for $R(P)$ imply \cite{Bergshoeff:2001hc}
\begin{eqnarray}
{\cal R}_{\mu \nu } = {\cal R}_{\nu \mu }\,, \qquad e_{[\m}{}^a \widehat
R_{\n\r ]}(D) = \widehat R_{[\m\n\r ]}{}^a(M)\,,\qquad
\widehat R_{\m\n}(D) = 0  \,. \label{Ricci}
\label{bi1}
\end{eqnarray}
The full commutator of two supersymmetry transformations is
\begin{eqnarray}
 \left[\d_Q(\e_1),\d_Q(\e_2)\right] &=&  \d_{cgct}(\xi_3^\m)+
\d_M(\l^{ab}_3) + \d_S(\eta_3)  + \d_U(\l^{ij}_3) +\d_K(\L^a_{K3}) \,, \label{algebraQQ}
\end{eqnarray}
where $\delta_{cgct}$ represents a covariant general coordinate transformation\footnote{The covariant general coordinate transformations are defined as $\delta_{cgct}(\xi) = \delta_{gct}(\xi) - \delta_{I}(\xi^{\m}h_{\m}^{I})$, where the index $I$ runs over all transformations except the general coordinate transformations and the $h_{\m}^{I}$ represent the corresponding gauge fields.}. The parameters appearing in (\ref{algebraQQ})
are
 \bea
\xi^\m_3       &=& \ft 12 \bar\e_2 \g^\m \e_1 \,,\nn\\
\l^{ab}_3      &=& - \rmi \bar \epsilon _2\gamma ^{[a}\gamma \cdot T
\gamma ^{b]}\epsilon _1 \,, \nonumber\\
\lambda^{ij}_3 &=& \rmi \bar\e^{(i}_2 \g\cdot T \e^{j)}_1\,, \nn\\
\eta^i_3       &=& - \ft {9}{4}\rmi\, \bar \e_2 \e_1 \chi^i
                   +\ft {7}{4}\rmi\, \bar \e_2 \g_c \e_1 \g^c \chi ^i \nn\\
               &&  + \ft1{4}\rmi\,  \bar \e_2^{(i} \g_{cd} \e_1^{j)}
\left( \g^{cd} \chi_j
                   + \ft 14\, \widehat R^{cd}{}_j(Q) \right) \, ,\nonumber\\
\Lambda^a_{K3} &=& -\ft 12 \bar\e_2\g^a\e_1 D + \ft{1}{96}
\bar\e^i_2\g^{abc}\e^j_1 \widehat R_{bcij}(V) \nn\\
               && + \ft1{12}\rmi\bar\e_2\left(-5\g^{abcd} D_b T_{cd} +
9 D_b T^{ba} \right)\e_1  \nn\\
               && + \bar\e_2\left(  \g^{abcde}T_{bc}T_{de}
                  - 4 \g^c T_{cd} T^{ad} +  \ft 23  \g^a T^2
                  \right)\e_1 \,.
 \eea

For the $Q,S$ commutators we find the following algebra
 \bea
 \left[\d_S(\eta),
\d_Q(\e)\right] &=& \d_D( \ft12\rmi \bar\e\eta ) + \d_M( \ft12\rmi \bar\e
\g^{ab} \eta) +      \d_U(  -\ft32\rmi \bar\e^{(i} \eta^{j)} )  +
\delta_K(\L_{3K}^a ) \,,\nn\\
\left[ \d_S(\eta_1), \d_S(\eta_2) \right] &=& \d_K( \ft 12 \bar\eta_2 \g^a
\eta_1 ) \,, \eea
 with
 \be
 \L_{3K}^a= \ft16 \bar{\e} \left(\g \cdot T \g^a - \ft12 \g^a \g \cdot T
\right) \eta \,. \label{algebraQSS}
\end{equation}
This concludes our review of the Standard Weyl Multiplet.
\subsection{The Vector Multiplet} \label{ss: vector}
The off-shell abelian $D=5$, $\mathcal{N}=2$ Vector Multiplet
contains $8+8$ degrees of freedom and consists of the fields
\begin{equation}
\{C_{\mu},\sigma,Y^{ij},\p^i\}\,,
\end{equation}
with Weyl weights (0,1,2,3/2) respectively. The bosonic field content
consists of a vector field $C_\m$, a scalar field $\sigma$ and
an auxiliary field $Y^{ij} = Y^{(ij)}$, that is an
$\SU(2)$ triplet. The fermion field is given by an $\SU(2)$ doublet
$\p^i$. 

The $Q$- and $S$-transformations of the Vector Multiplet, in the
background of the Standard Weyl Multiplet, are given by \cite{Bergshoeff:2002qk}
\begin{eqnarray}
\d C_\m &=& -\ft12\rmi \s \bar{\e} \p_\m + \ft12 \bar{\e}
\g_\m \p \ ,
\nn\\
\d Y^{ij} &=& -\ft12
\bar{\e}^{(i} \slashed{\mathcal{D}} \p^{j)} + \ft12 \rmi \bar{\e}^{(i}
\g \cdot T \p^{j)} - 4 \rmi \s \bar{\e}^{(i} \chi^{j)} +
\ft12 \rmi \bar{\eta}^{(i} \p^{j)}\,, \nn\\
\d\p^{i} &=& - \ft14 \g \cdot \widehat{G} \e^i -\ft12\rmi
\slashed{\mathcal{D}}\s \e^i + \s \g \cdot T \e^i - Y^{ij} \e_j +
\s \eta^i \ ,
\nn\\
\d\s &=& \ft12 \rmi \bar{\e}
\p . \label{trvm}
\end{eqnarray}
We have used here the superconformally covariant derivatives
\begin{eqnarray}
\mathcal{D}_\mu\, \s &=&  (\partial_\mu - b_\mu) \s
- \tfrac12\, \rmi\bar{\psi}_\mu \p \ ,
\nn\\
\mathcal{D}_\mu \p^i &=&  (\partial_\mu -\ft32 b_\mu +\ft14\,  \o_\mu{}^{ab}\g_{ab} ) \p^{i} - V_\mu^{ij} \p_j \nn\\
&& +\tfrac 14 \g
\cdot \widehat{G} \p_\mu^i + \ft12\rmi \slashed{\mathcal{D}} \s \p_\mu^i
 + Y^{ij} \p_{\mu\, j} - \s  \g \cdot T \p_\mu^i - \s
\phi_\mu^i,
\label{cd2}
\end{eqnarray}
and the supercovariant Yang-Mills curvature
\begin{equation}
\widehat{G}_{\m\n}  = G_{\m\n} - \bar{\p}_{[\m} \g_{\n]} \p + \tfrac 12 \rmi
\s \bar{\p}_{[\m} \p_{\n]}\,,
\label{hatF}
\end{equation}
where $G_{\m\n}=2 \partial_{[\mu } C_{\nu ]}$. The supersymmetry transformation rule for $\widehat{G}_{\m\n}$ is given by
\bea
\d \widehat{G}_{\m\n} = - \tfrac12 \rmi \s \bar\e \widehat{R}_{\m\n}(Q) - \bar\e \g_{[\m} {\cal{D}}_{\n ]}\p + \rmi \bar\e \g_{[\m} \g \cdot T \g_{\n ]} \p + \rmi \bar\eta \g_{\m\n} \p\,.
\label{TransG}
\eea
This concludes our discussion on the Vector Multiplet.
\subsection{The Linear Multiplet} \label{ss: linear}
The off-shell $D=5, {\cal{N}} = 2$ Linear Multiplet contains $8+8$ degrees of freedom and consists of the fields
\begin{equation}
\{L^{ij},E^a,N,\varphi^i\}\,,
\end{equation}
with Weyl weights (3,4,4,7/2) respectively. The bosonic fields consist of an $\SU(2)$ triplet $L^{ij} = L^{(ij)}$, a constrained vector $E_{a}$ and a scalar $N$. The fermion field is given by an $\SU(2)$ doublet
$\varphi^i$. The $Q$ and $S$ supersymmetry transformations of the Linear Multiplet in the background of the Standard Weyl Multiplet are given by
\begin{eqnarray}
\delta L^{ij} &=& i\bar{\e}^{(i}\varphi^{j)}\,,\nn\\
\delta \varphi^{i} &=& - \tfrac{1}{2} i \slashed{\mathcal{D}}L^{ij}\e_{j} - \tfrac{1}{2} i \gamma^{a} E_{a} \e^i + \tfrac{1}{2} N \e^{i}  - \g \cdot T L^{ij} \e_{j} + 3 L^{ij}\eta_{j}\,,\nn\\
\delta E_{a} &=& -\tfrac{1}{2} i \bar{\e} \g_{ab} \mathcal{D}^{b} \varphi  - 2 \bar{\e} \gamma^{b} \varphi T_{ba} - 2\bar{\eta} \g_{a} \varphi\,, \nn\\
\delta N &=& \tfrac{1}{2} \bar{\e} \slashed{\mathcal{D}}\varphi  + \tfrac{3}{2} i \bar{\e} \g \cdot T \vf + 4 i \bar{\e}^i \chi^j L_{ij} + \tfrac{3}{2} i \bar{\eta} \varphi\,,
\label{trlm}
\end{eqnarray}
where we used the following superconformally covariant derivatives
\bea
\mathcal{D}_\m L^{ij}&=&(\partial_{\m}-3b_{\m})L^{ij}+2V_{\m}{}^{(i}{}_kL^{j)k}-i\bar{\p}_{\m}^{(i}\varphi^{j)}\,, \nn \\
\mathcal{D}_\m \varphi^i&=&(\partial_{\m}-\tfrac{7}{2}b_\m +\ft14\o_\m{}^{ab}\g_{ab})\varphi^i-V_{\m}^{ij}\varphi_j+\tfrac{1}{2} i \slashed{\mathcal{D}}L^{ij}\p_{\m\,j} + \tfrac{1}{2} i \gamma^{a} E_{a} \p_\m^i \nn \\
&&- \tfrac{1}{2} N \p_\m^{i}  + \g \cdot T L^{ij} \p_{\m\,j} - 3 L^{ij}\phi_{\m\,j}\,, \nn \\
\mathcal{D}_\m E_a &=&(\partial_\m-4b_\m)E_a+\o_{\m ab}E^b+\tfrac{1}{2} i \bar{\p}_\m \g_{ab} \mathcal{D}^{b} \varphi  +2 \bar{\p}_\m \gamma^{b} \varphi T_{ba} + 2\bar{\phi}_\m \g_{a} \varphi\,.
\eea
Finally, we note that the superconformal algebra closes if the following constraint is satisfied
\begin{eqnarray}
\mathcal{D}^{a} E_{a}= 0\,. \label{closure}
\end{eqnarray}
The solution for $E_a$ in terms of a 3-form $E_{\m\n\r}$ is
\bea
E^a = - \tfrac{1}{12}e_{\m}{}^{a}e^{-1}\varepsilon^{\m\n\r\s\l}\mathcal{D}_\n  E_{\r\s\l}\,
\label{3formE}
\eea
and $E_{\m\n\r}$ has the following gauge invariance
\bea
\d_\L E_{\m\n\r} = 3 \partial_{[\m} \L_{\n\r ]}\,.
\label{GInvE}
\eea
Also, for the dual 2-form field we have
\bea
E^a &=& e_\m{}^a {\cal{D}}_\n E^{\m\n}\,, \nn\\
E_{\m\n\r} &=& e\varepsilon_{\m\n\r\s\l} E^{\s\l}\,, \nn\\
\d E^{\m\n} &=& - \tfrac12 \rmi \bar\e \g^{\m\n} \vf - \tfrac12 \bar\p_\r^i \g^{\m\n\r} \e^j L_{ij} - \partial_\r \tilde{\L}^{\m\n\r}\,, \nn\\
{\cal{D}}_\n E^{\m\n} &=& \partial_\n E^{\m\n} + \tfrac12 \rmi \bar\p_\n \g^{\m\n} \vf + \tfrac14 \bar\p_\r^i \g^{\m\n\r} \p_\n^j L_{ij}\,.
\label{dualityforE}
\eea
This concludes our discussion on the Linear Multiplet.


\section{Actions}\label{section: actions}
In this section, we will construct the action for a Linear Multiplet and present the action for Vector Multiplet coupled to the Standard Weyl Multiplet \cite{Bergshoeff:2002qk}. Our starting point is a density formula for the product of a Vector Multiplet and a Linear Multiplet. This will be presented in subsection \ref{ss: density formula}. In subsection \ref{ss: action linear} we will use this formula, after embedding the Linear Multiplet into a Vector Multiplet, to construct the superconformal action for the Linear Multiplet. In the last subsection \ref{ss: action vector} we present the action for the Vector Multiplet.

\subsection{Density formula} \label{ss: density formula}
We need an expression constructed from the components of the Linear and Vector Multiplet that can be used as a superconformal action. In \cite{Kugo:2000hn} a density formula is given for the product of a Vector Multiplet and a Linear Multiplet
\begin{eqnarray}
e^{-1}{\cal{L}}_{VL} &=& Y^{ij}L_{ij} + i \Bar{\p} \vf - \tfrac{1}{2} \bar{\p}_{a}^{i} \g^{a} \p^{j}L_{ij} + C_{a} P^{a} \nn\\
&& + \sigma (N + \tfrac{1}{2} \bar{\p}_{a}\g^{a}\vf + \tfrac{1}{4}i \bar{\p}_{a}^{i} \g^{ab} \p_{b}^{j} L_{ij} ) \,,
\label{VLaction1}
\end{eqnarray}
where $P_\m$, the pure bosonic part of the supercovariant field $E_\m$, is defined as
\begin{equation}
P^{a}=E^{a} + \tfrac{1}{2} i \bar{\p}_{b} \g^{ba} \vf + \tfrac{1}{4} \bar{\p}_{b}^{i} \g^{abc} \p_{c}^{j}L_{ij}\,.
\label{Pa}
\end{equation} 
Using (\ref{3formE}), we can express $P^a$ as
\bea
P^a = -\tfrac{1}{12}e_\m{}^a e^{-1} \varepsilon^{\m\n\r\s\l} \partial_\n E_{\r\s\l}\,.
\label{defP}
\eea
Using (\ref{defP}) and (\ref{dualityforE}), one can rewrite  ${\cal{L}}_{VL}$ as
\bea
e^{-1}{\cal{L}}_{VL} &=& Y^{ij}L_{ij} + i \bar{\p} \vf - \tfrac{1}{2} \bar{\p}_{a}^{i} \g^{a} \p^{j}L_{ij} +\tfrac{1}{2} G_{\m\n} E^{\m\n} \nn\\
&& + \sigma (N + \tfrac{1}{2} \bar{\p}_{a}\g^{a}\vf + \tfrac{1}{4}i \bar{\p}_{a}^{i} \g^{ab} \p_{b}^{j} L_{ij} )\,. \label{VLaction2}
\eea

\subsection{Action for the Linear Multiplet} \label{ss: action linear}
We want to use the density formula (\ref{VLaction1}) to construct an action for the Linear Multiplet. Hence, we start with embedding the components of the Linear Multiplet $(L_{ij}, \vf^i, E_a, N)$ into the components of the Vector Multiplet $(Y^{ij},C_\m, \s, \p^i)$. Such embeddings are already considered in 4 and 6 dimensions \cite{Coomans:2011ih, Bergshoeff:1985mz, deWit:1982na} and here we will follow the same procedure.

As described in \ref{ss: linear} the  Linear Multiplet consists of a triplet of scalars $L_{ij}$, a constrained vector $E_{a}$, a doublet of Majorana spinors $\vf^i$ and a scalar $N$. One starts the construction of the Vector Multiplet with the identification $\s = N$, where $\s$ is the scalar of the Vector Multiplet. There is, however, a mismatch between the Weyl weights of these fields. Therefore one needs another scalar field to compensate for this mismatch. For this we will use
\bea
L ^2 = L_{ij} L^{ij}\,.
\label{scalardef}
\eea
We can then identify the scalar of the Vector Multiplet as  $ \s = 2 L^{-1} N + \rmi \bar{\vf}_i \vf_j L^{ij} L^{-3}$. This identification has the correct Weyl weight, and the second term is the supersymmetric completion that is determined by the $S$-invariance of $\s$. Upon applying a sequence of supersymmetry transformations, we obtained the following embeddings
\bea
\s &=& 2 L^{-1} N + (\varphi^i\text{-terms})\,, \nn\\
\p_i &=& -2 \rmi \slashed{\cal{D}}\vf_i L^{-1} + 16L^{-1}L_{ij}\chi^j + (\varphi^i\text{-terms})\,, \nn \\
Y_{ij} &=& L^{-1} \Box^C L_{ij} - {\cal{D}}_{a}L_{k(i} {\cal{D}}^{a}L_{j)m} L^{km} L^{-3} - N^2 L_{ij} L^{-3} - E_{\m}E^{\m}L_{ij} L^{-3}\nn\\ 
&& + \tfrac{8}{3} L^{-1} T^2 L_{ij} + 4 L^{-1} D L_{ij} + 2 E_{\m} L_{k(i} {\cal{D}}^{\m} L_{j)}{}^{k} L^{-3} + (\varphi^i\text{-terms})\,, \nn\\
\widehat{G}_{\m\n} &=& 4 {\cal{D}}_{[\m}(L^{-1} E_{\n ]}) + 2L^{-1} \widehat{R}_{\m\n}{}^{ij}(V) L_{ij} - 2 L^{-3} L_{k}^{l} {\cal{D}}_{[\m}L^{kp} {\cal{D}}_{\n ]} L_{lp} + (\varphi^i\text{-terms})\,.
\label{embed}
\eea
Here, we did not write the fermionic terms proportional to $\vf^i$ explicitly because in the following section we will set $\varphi^i$ to zero to fix the $S$-gauge. 

After plugging the components (\ref{embed}) into the density formula (\ref{VLaction1}) we end up with a superconformal action for the Linear Multiplet
\bea
e^{-1}{\cal{L}}_L &=& L^{-1} L_{ij} \Box^c L^{ij} - L^{ij} {\cal{D}}_{\m}L_{k(i} {\cal{D}}^{\m} L_{j)m} L^{km} L^{-3} - N^2 L^{-1} - E_{\m} E^{\m} L^{-1} \nn\\
&& + \tfrac{8}{3} L T^{2} + 4 D L - \tfrac{1}{2}L^{-3} E^{\m\n} L_{k}^{l} \partial_{\m} L^{kp} \partial_\n L_{pl} + 2 E^{\m\n} \partial_{\m} ( L^{-1} E_{\n} + V_{\n}^{ij} L_{ij} L^{-1} ) \nn\\
&& + \rmi L^{-1}L_{ij} \bar{\p}_{\m}^i\g^{\m} \slashed{\cal{D}}\vf^j + 4 L \bar{\p}_{\m} \g^{\m} \chi +\tfrac{1}{2}iL^{-1}N\bar{\p}_\m^i\g^{\m\n}\p_\n^jL_{ij}+(\varphi^i\text{-terms})\,,
\label{SCAction}
\eea
where the superconformal d'Alembertian is given by
\bea
L_{ij} \Box^c L^{ij}&=&L_{ij}(\partial^a-4b^a+\omega_b{}^{ba}){\cal{D}}_a L^{ij}+2L_{ij} V_a{}^i{}_k{\cal{D}}^a L^{jk}+6L^2f_a{}^a \nn \\
&&-iL_{ij}\bar{\psi}^{a i}{\cal D}_a\varphi^j-6L^2\bar{\psi}^a\g_a\chi-L_{ij}\bar{\varphi}^i\g\cdot T\g^a\psi_a^j+L_{ij}\bar{\varphi}^i\g^a\phi_a^j\,.
\eea

\subsection{Action for the Vector Multiplet} \label{ss: action vector}

A Vector Multiplet can be embedded into the Linear Multiplet to construct an action using the invariant action formula \ref{VLaction1}. The action for the abelian Vector Multiplet up to 4-fermion
terms reads \cite{Bergshoeff:2002qk}
\bea
e^{-1}{\cal{L}}_V &=& - \tfrac14 \s G_{\m\n}G^{\m\n} + \tfrac13 \s^2 \Box^c \s + \tfrac16 \s {\cal{D}}_\m \s {\cal{D}}^\m \s + \s Y_{ij} Y^{ij} \nn\\
&& - \tfrac{4}{3} \s^3 \Big(  D + \tfrac{26}3 T^2 \Big) + 4 \s^2 G_{\m\n} T^{\m\n} - \tfrac1{24} e^{-1}\varepsilon^{\m\n\r\s\l} C_\m G_{\n\r} G_{\s\l} \nn\\
&& - \tfrac{1}{2}\s \bar{\p} \slashed{\cal{D}} \p  - \tfrac 18 \rmi \bar{\p} \g
\cdot G \p  - \tfrac 12
\rmi \bar{\p}^{i} \p^j Y_{ij}  + \rmi \s  \bar\p
\g \cdot T \p - 8 \rmi \s^2 \bar\p \chi \nn \\
&&+\tfrac 16 \s \bar \p_\m \g^\m \!\left(\!\rmi\s \slashed{\cal{D}} \p + \tfrac 12 \rmi \slashed{\cal{D}} \s  \p  - \tfrac 14 \g
{\cdot} G  \p
+ 2 \s \g {\cdot} T \p  - 8 \s^2 \chi \!\right)\! \nn\\
&&\hphantom{\! \biggl[}
{-} \tfrac 16 \bar \p_\m \g_\n \p \left(-2\s G^{\m\n} -8 \s^2 T^{\m\n} \right) -\tfrac 1{12} \s \bar \p_\l \g^{\m\n\l} \p G_{\m\n}  \nn\\
&&\hphantom{\! \biggl[} {+}\tfrac 1{12} \rmi \s \bar \p_\m
\p_\n  \left(-2\s G^{\m\n} -8
\s^2  T^{\m\n} \right) +\tfrac 1{48} \rmi
\s^2  \bar \p_\l \g^{\m\n\l\r} \p_\r
G_{\m\n}  \nn \\&& \hphantom{\!
\biggl[} {-} \tfrac 12 \s \bar \p_\m^i \g^\m \p^{j} Y_{ij} +\tfrac 1{6} \rmi \s^2 \bar\p_\m^i \g^{\m\n} \p_\n^j Y_{ij}\,,
\label{vectoractionf}
\eea
where
\bea
\Box^c \s &=& (\partial^a - 2 b^a + \o_b{}^{ba} ){\cal{D}}_a \s  - \tfrac12 \rmi \bar\p_a {\cal{D}}^a \p - 2\s \bar\p_a \g^a \chi \nn\\
&& + \tfrac12 \bar\p_a \g^a \g \cdot T \p + \tfrac12 \bar\phi_a \g^a \p + 2 f_a^a \s\,.
\label{Box1}
\eea

\section{Ungauged \texorpdfstring{$D = 5, \, {\cal{N}} = 2$}{} Supergravity}\label{ungaugedtheory}

\begin{figure}
\caption{The schematic description of how to obtain off-shell ${\cal{N}}=2, D=5$ Poincar\'e Supergravity}
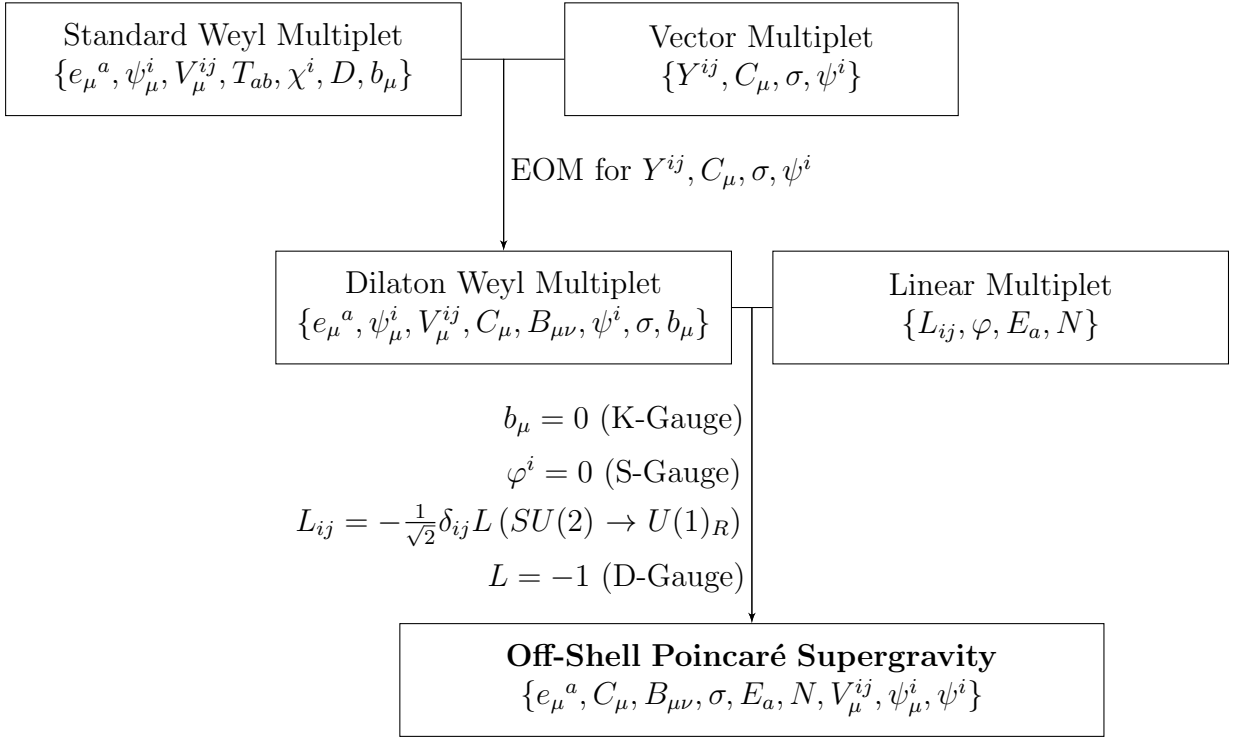
\tikzstyle{decision} = [diamond, draw]
\tikzstyle{line} = [draw, -latex']
\tikzstyle{block} = [draw,rectangle,text width=14em, text centered,minimum height=15mm, node distance=10em]
\begin{tikzpicture}
\node [block, text width=12em]  (Vector) {Vector Multiplet\\ \{$Y^{ij}, C_\m, \s, \p^i$\}};
\node [block, left of=Vector, xshift=-7em] (Standard) {Standard Weyl Multiplet\\ \{$e_{\m}{}^{a}, \p_\m^i, V_\m^{ij}, T_{ab}, \chi^i, D, b_\m$\}};
\node [block, below of=Vector, xshift=-8.3em,yshift=2em] (Dilaton) {Dilaton Weyl Multiplet\\ \{$e_{\m}{}^{a}, \p_\m^i, V_\m^{ij}, C_\m, B_{\m\n}, \p^i, \s, b_\m$\}};
\path[line]  (Vector) -| node[yshift=-1.5cm, xshift=2.1cm] {EOM for $Y^{ij}, C_\m, \s, \p^i$} (Dilaton) ;
\path[line]  (Standard) -| (Dilaton);
\node [block, right of=Dilaton, xshift=6em] (Linear) {Linear Multiplet\\ \{$L_{ij}, \vf, E_a, N$\}};
\node [block, below of=Dilaton,yshift=-2em,,text width=22em,xshift=3.3cm] (PoincareMatter) {\textbf{Off-Shell Poincar\'e Supergravity} \\ \{$e_\m{}^a, C_\m, B_{\m\n}, \s, E_a, N, V_\m^{ij}, \p_\m^i,\p^i$\}};
\path[line]  (Dilaton) -| node[yshift=-1.5cm, xshift=-1.7cm] {$b_\m=0$ (K-Gauge)} node[yshift=-2.2cm, xshift=-1.7cm] {$\vf^i=0$ (S-Gauge)}node[yshift=-2.9cm, xshift=-3.1cm] {$L_{ij}=-\frac{1}{\sqrt{2}} \d_{ij}L$\,($SU(2)$ $\rightarrow$ $U(1)_R$) } node[yshift=-3.6cm, xshift=-1.8cm] {$L=-1$ (D-Gauge)}(PoincareMatter) ;
\path[line]  (Linear) -| (PoincareMatter);
\end{tikzpicture}
\label{fig;1}
\end{figure}

In the previous section we have constructed an action for the Linear Multiplet and presented the action for the Vector Multiplet constructed in \cite{Bergshoeff:2002qk}. A naive attempt to construct the Poincar\'e supergravity directly by gauge fixing the Linear or Vector Multiplet action obviously fails as the equation of motion for $D$ is inconsistent. In this section we combine both actions to obtain a theory in a Dilaton Weyl background \cite{Bergshoeff:2001hc} which will solve the inconsistency for $D$. In the first subsection \ref{ss: construction} we combine the Linear Multiplet action and the Vector Multiplet action and solve for Standard Weyl matter fields. The resulting theory is the superconformal theory with the matter fields of the Dilaton Weyl Multiplet.  In the next subsection \ref{ss: gaugefix}, we discuss the procedure to gauge fix the redundant conformal symmetries, and give the pure off-shell Poincar\'e Supergravity. The schematic description of how to obtain this theory is given in figure \ref{fig;1}.

\subsection{Construction of the Superconformal Action} \label{ss: construction}
Our starting point is the following Lagrangian
\begin{equation}
\mathcal{L}_0=\mathcal{L}_L-3\mathcal{L}_V\,, \label{VplusL}
\end{equation}
where $\mathcal{L}_L$ and $\mathcal{L}_V$ are given in (\ref{SCAction}) and (\ref{vectoractionf}), and the factor of 3 in front of the Vector Multiplet action is chosen for later convenience. The equation of motion for the auxiliary $Y^{ij}$, and the field equations for $\sigma$ and $\p^i$ allow us to express the Standard Weyl matter fields $Y^{ij}, D$ and $\chi^i$ in terms of the fields of the Vector Multiplet \cite{Bergshoeff:2001hc}
\bea
Y^{ij}&=&\tfrac{1}{4} \rmi \s^{-1} \bar\p^i \p^j\,,
\eea
\bea
\chi^i &=& \tfrac18 \rmi \s^{-1}\slashed{\cal{D}}\p^i + \tfrac1{16} \rmi \s^{-2}\slashed{\cal{D}} \s \p^i - \tfrac{1}{32} \s^{-2} \g \cdot G \p^i \nn\\
&& + \tfrac14 \s^{-1} \g \cdot T \p^i + \tfrac1{8}\s^{-2}Y^{ij}\p_j\,, \\
D &=& \tfrac14 \s^{-1} \Box^c \s +\tfrac{1}{8} \s^{-2}( {\cal{D}}_a \s)( {\cal{D}}^a \s) - \tfrac1{16} \s^{-2} G_{\m\n} G^{\m\n} \nn\\
&& - \tfrac{1}{8} \s^{-2} \bar\p{\slashed{\cal{D}}} \p +\tfrac{1}{4}\sigma^{-2}Y^{ij}Y_{ij}- 4\rmi \s^{-1} \bar\p \chi +\tfrac{1}{8}\sigma^{-2}\opsi_{\m}\g_{\n}\p G^{\m\n} \nn \\
&&-\tfrac{1}{16}i\sigma^{-1}\opsi_{\m}\p_{\m}G^{\m\n} + \Big( -\tfrac{26}3 T_{ab} + 2 \s^{-1} \widehat G_{ab} + \tfrac14 \rmi \s^{-2} \bar\p \g_{ab} \p \Big) T^{ab}\,,
\label{IdenYChiD}
\eea
where $\chi^i$ and $D$ are given up to 3- and 4-fermion terms respectively. The equation of motion for $C_\m$ implies a Bianchi identity for an antisymmetric 2-form tensor gauge field $B_{\m\n}$
\bea
{\cal{D}}_{[a} \widehat H_{bcd]} &=& \tfrac34\widehat G_{[ab} \widehat G_{cd]}\,,
\label{BianchiB}
\eea
where the 3-form curvature tensor $ \widehat H_{abc}$ is defined as
\bea
\tfrac16 \varepsilon_{abcde} \widehat H^{cde} &=& 8 \s^2 T_{ab} - \s  \widehat G_{ab} - \tfrac14 \rmi \bar\p \g_{ab} \p\,.
\label{IdenT}
\eea
This equation allows us to identify $T_{ab}$ in terms of the elements of the Vector Multiplet and the supercovariant field strength $ \widehat H_{\m\n\r}$. As a matter of fact, the fields $\s, C_\m, B_{\m\n}$ and $\p^i$ along with the gauge fields $e_\m{}^a, \p_\m^i, b_\m, V_{\m}{}^{ij}$ form an alternative Weyl multiplet: the Dilaton Weyl Multiplet. Equations (\ref{IdenYChiD}) and (\ref{IdenT}) present a map between the Standard Weyl Multiplet and the Dilaton Weyl Multiplet.

The $Q$, $S$ and $K$ supersymmetry transformations for the fields of the Dilaton Weyl Multiplet are given by (up to 3-fermion terms)
\bea
\d e_\m{}^a   &=&  \ft 12\bar\e \g^a \psi_\m  \nn\, ,\\
\d \psi_\m^i   &=& (\partial_\m+\tfrac{1}{2}b_\m+\tfrac{1}{4}\omega_\m{}^{ab}\g_{ab})\e^i-V_\m^{ij}\e_j + \rmi \g\cdot \underline{T} \g_\m
\e^i - \rmi \g_\m
\eta^i  \nn\, ,\\
\d V_\m{}^{ij} &=&  -\ft32\rmi \bar\e^{(i} \phi_\m^{j)} +4
\bar\e^{(i}\g_\m \underline{\chi}^{j)}
  + \rmi \bar\e^{(i} \g\cdot \underline{T} \psi_\m^{j)} + \ft32\rmi
\bar\eta^{(i}\psi_\m^{j)} \nn\, ,\\
 \d C_\m
&=& -\ft12\rmi \s \bar{\e} \p_\m + \ft12
\bar{\e} \g_\m \p, \nonumber\\
 \d B_{\m\n}
&=& \ft12 \s^2 \bar{\e} \g_{[\m} \p_{\n]} + \ft12 \rmi \s \bar{\e}
\g_{\m\n} \p + C_{[\m} \d(\e) C_{\n]}, \nonumber\\
\d \p^i &=& - \ft14 \g \cdot \widehat{G} \e^i -\ft12\rmi \slashed{\mathcal{D}} \s
\e^i + \s \g \cdot \underline{T} \e^i -\underline{Y}^{ij}\e_j + \s\eta^i \,,\nonumber\\
\d \s &=& \ft12 \rmi \bar{\e} \p \, ,\nonumber\\
 \d b_\m       &=& \ft12 \rmi \bar\e\phi_\m -2 \bar\e\g_\mu \underline{\chi} +
\ft12\rmi \bar\eta\psi_\mu+2\Lambda _{K\mu } \,,
\label{TransDW}
\eea
where we have underlined\footnote{The expression for $Y^{ij}$ in (\ref{IdenYChiD}) is bilinear in the fermions and gives rise to 3-fermion terms in the transformation rule of $\p^i$. However, we did include the term in $\d \p^i$ because, when we will discuss the gauged theory in the next section, the expression for $Y^{ij}$ gets deformed by a purely bosonic term.} $Y^{ij}$, $T_{\m\n}$, $D$ and $\chi^i$ to indicate that they are now composite fields with the definitions given in  (\ref{IdenT}) and (\ref{IdenYChiD}) respectively. The commutator of Q-transformations picks up the following modifications
\bea
[\d(\e_1), \d(\e_2) ] &=& \ldots + \d_{U(1)} (\L_3=- \tfrac12 \rmi \s\bar\e_2 \e_1) + \d_B (-\tfrac14 \s^2 \bar\e_2 \g_\m \e_1 - \tfrac12 C_\m \L_3)\,,
\label{ModifQComm}
\eea
where the ellipses refer to the standard commutation rule and $\d_B$ is a vector gauge transformation for the field $B_{\m\n}$. From the transformation rule (\ref{TransDW}) for $B_{\m\n}$, we identify the supercovariant field strength $\widehat H_{\m\n\r}$ as
\bea
\widehat{H}_{\m\n\r} &=&3\partial _{[\mu }B_{\nu \rho ]} + \ft32 C_{[\m}  G_{\n\r]} - \ft34
\s^2 \bar{\p}_{[\m} \g_\n \p_{\r]} - \ft32\rmi \s \bar{\p}_{[\m}
\g_{\n\r]} \p\,
\label{DefH}
\eea
and the supersymmetry transformation rule for $\widehat{H}_{\m\n\r}$ is given by
\bea
\d \widehat{H}_{abc} &=& -\ft34 \s^2 \bar\e \g_{[a} \widehat{R}_{bc]}(Q) + \ft32 \rmi \bar\e \g_{[ab} {\mathcal{D}}_{c]} \p + \ft32 \rmi \bar\e \g_{[ab} \p {\mathcal{D}}_{c]} \s\nn\\
&& -\ft32 \s \bar\e \g_{[a} \g \cdot T \g_{bc]} \p -\ft32 \bar\e \g_{[a} \widehat{G}_{bc]} \p - \ft32 \s \bar\eta \g_{abc} \p\,.
\eea
For $\widehat{H}_{\m\n\r}$ to be gauge invariant $B_{\m\n}$ should transform under gauge transformations as follows 
\bea
\d_\L B_{\m\n} = 2 \partial_{[\m} \L_{\n ]} - \tfrac12 \L G_{\m\n}\,.
\eea
Plugging in the expressions for $Y^{ij}$ and the Standard Weyl matter fields, (\ref{IdenYChiD}) and (\ref{IdenT}), into the Lagrangian (\ref{VplusL}), we obtain the following superconformal action
\bea
e^{-1}{\cal{L}}_{0,DW}&=& L^{-1} L_{ij} \Box^C L^{ij} - L^{ij} {\cal{D}}_{\m}L_{ki} {\cal{D}}^{\m} L_{jm} L^{km} L^{-3} - N^2 L^{-1} - E_{\m} E^{\m} L^{-1} \nn\\
&& + \tfrac{8}{3} L \underline{T}^{2} + 4\underline{D} L - \tfrac{1}{2}L^{-3} E^{\m\n} L_{k}^{l} \partial_{\m} L^{kp} \partial_\n L_{pl} + 2 E^{\m\n} \partial_{\m} ( L^{-1} E_{\n} + V_{\n}^{ij} L_{ij} L^{-1} ) \nn\\
&& + \rmi L^{-1}L_{ij} \bar{\p}_{\m}^i\g^{\m} \slashed{\cal{D}}\vf^j + 4 L \bar{\p}_{\m} \g^{\m} \underline{\chi} +\tfrac{1}{2}iL^{-1}N\bar{\p}_\m^i\g^{\m\n}\p_\n^jL_{ij}+(\varphi^i\text{-terms})\,,
\label{FullSCAction}
\eea
where the subscript $DW$ indicates the fact that this Lagrangian is in the background of the Dilaton Weyl Multiplet. In the following subsection we will gauge fix  the redundant conformal symmetries to obtain off-shell pure $D = 5, \, {\cal{N}}=2$ supergravity.

\subsection{Gauge Fixing, Decomposition Rules and the Off-Shell Poincar\'e Action}  \label{ss: gaugefix}
The gauge fixing conditions we adopt here are
\bea
L_{ij} = -\tfrac{1}{\sqrt{2}} \d_{ij}, \qquad b_{\m} = 0, \qquad \vf^i = 0\,.
\label{gaugefix}
\eea
The first gauge condition breaks the $SU(2)$ symmetry down to $U(1)_{R}$, and fixes the dilatation symmetry by the choice $L=-1$. The second condition fixes the special conformal symmetry whereas the last one fixes the S-supersymmetry.\\
In order for these gauge conditions to be invariant under supersymmetry, one needs to add compensating conformal boost transformations with parameter
\bea
\L_{K\m} = -\tfrac{1}{4} \rmi \bar{\e}\phi_{\m} - \tfrac{1}{4} \rmi \bar{\eta}\p_{\m} + \bar{\e}\g_\m \chi\,
\label{decomK}
\eea
and compensating S-supersymmetry transformations with parameter
\bea
\eta_k = \tfrac{1}{3} \Big(  \g \cdot T \e_k + \tfrac1{\sqrt{2}} N \d_{ik} \e^i -  \tfrac1{\sqrt{2}} \rmi \slashed{E} \d_{ik} \e^i -   \rmi\g^\m V_{\m}{}^{(i}{}_{l} \d^{j)l}\d_{ik}\e_j \Big)\,. \label{decomS}
\eea
Performing all the steps of gauge fixing, and using the expressions for the dependent gauge fields (\ref{transfDepF}) into the Lagrangian (\ref{FullSCAction}), we end up with the following off-shell Lagrangian for ${\cal{N}} = 2, \, D=5$ Poincar\'e supergravity (up to 4-fermion terms)
\bea
e^{-1}{\cal{L}}_{0,DW}|_{L=-1}&=& \tfrac{1}{2} R - \tfrac{1}{4} \s^{-2} G_{\m\n} G^{\m\n} - \tfrac{1}{6} \s^{-4} H_{\m\n\r} H^{\m\n\r} - \tfrac{3}{2} \s^{-2} \partial_\m \s \partial^\m \s \nn\\
&& - N^2 +  P_{\m} P^\m + \sqrt{2} P_\m V^\m - V'_{\m}{}^{ij}V'^{\m}_{ij} \nn \\
&&-\tfrac{1}{2} \bar{\p}_{\m} \g^{\m\n\r} D_\n \p_\r -\tfrac{3}{2}\sigma^{-2}\bar{\p}\slashed{D}' \p-\tfrac{3}{2}i\sigma^{-2}\bar{\p}\g^{\m}\g^{\r}\p_{\m}\partial_{\r}\sigma \nn \\
&&-\tfrac{1}{8}i\sigma^{-1}\bar{\p}_{\m}\g^{\m\n\r\s}\p_{\n}G_{\r\s}-\tfrac{1}{4}i\sigma^{-1}\bar{\p}_{\m}\p_{\n}G^{\m\n}-\tfrac{1}{4}\sigma^{-2}\bar{\p}_{\m}\g^{\m\n\r}\p G_{\n\r}  \nn \\
&&+\tfrac{1}{2}\sigma^{-2}\bar{\p}_{\m}\g_{\n}\p G^{\m\n}-\tfrac{1}{8}i\sigma^{-3}\bar{\p}\g\cdot G\p +\tfrac{1}{2\sqrt{2}}\bar\p_\m^i\g^{\m\n\r}\p_\n^j P_\r \d_{ij}\nn \\
&&-\tfrac{1}{24}\sigma^{-2}\bar{\p}_{\m}\g^{\m\n\r\s\l}\p_{\n}H_{\r\s\l}+\tfrac{1}{4}\sigma^{-2}\bar{\p}_{\m}\g_{\n}\p_{\r}H^{\m\n\r} \label{ActionBos} \\
&&-\tfrac{1}{6}i\sigma^{-3}\bar{\p}_{\m}\g^{\m\n\r\s}\p H_{\n\r\s}+\tfrac{1}{2}i\sigma^{-3}\bar{\p}_{\m}\g_{\n\r}\p H^{\m\n\r}-\tfrac{5}{24}\sigma^{-4}\bar{\p}\g\cdot H\p\,, \nn 
\eea
where the subscript $L=-1$ is shorthand for the gauge fixing described in (\ref{gaugefix}). Notice that we have decomposed the field $V_\m^{ij}$ into its trace and traceless part, i.e. $V_\m^{ij} = V_\m^{'ij} + \tfrac12 \d^{ij} V_\m$ with $V_\m^{'ij} \d_{ij}=0$. The 2- and 3-form field strengths are defined as
\bea
G_{\m\n} &=& 2 \partial_{[\m}C_{\n ]}\,,\nn\\
H_{\m\n\r} &=& B_{\m\n\r} + \tfrac32 C_{[\m} G_{\n\r ]}\,,
\label{GandH}
\eea
where $ B_{\m\n\r}=3\partial_{[\m}B_{\n\r ] }$, and the $U(1)_R$ covariant derivative $D_\m \p_\n^i$ and full $SU(2)$ covariant derivative $D'_\m \p^i$ are defined as
\bea
D_\m \p_\n^i&=&\Bigl(\partial_\m+\tfrac{1}{4}\omega_\m{}^{ab}\g_{ab}\Bigr)\p_\n^i\ -\tfrac12 V_\m \d^{ij} \p_{\n\, j}\,, \nn \\
D'_\m \p^i&=&\Bigl(\partial_\m+\tfrac{1}{4}\omega_\m{}^{ab}\g_{ab}\Bigr)\p^i+V'_\m{}^i{}_j\p^j-\tfrac12 V_\m \d^{ij} \p_{j}\,.
\label{covder}
\eea
The ${\cal{N}}=2$ off-shell supergravity that we constructed above by means of superconformal tensor calculus has the following field content
\bea
(e_\m{}^a,\, \p_\m^i,\, C_\m,\, B_{\m\n},\,\p^i,\,\s,\, E_\m ,\, N,\, V_\m ,\, V_\m^{'ij} )
\label{fieldcontent}
\eea
with $(10, 32, 4, 6, 8, 1, 4, 1, 4, 10)$ off-shell degrees of freedoms respectively. Therefore our off-shell Poincar\'e multiplet has 40 + 40 off-shell degrees of freedom. The supersymmetry transformations, up to 3-fermions, are
\bea
\d e_\m{}^a   &=&  \ft 12\bar\e \g^a \psi_\m  \nn\, ,\\
\d \psi_\m^i   &=& (\partial_\m+\tfrac{1}{4}\omega_\m{}^{ab}\g_{ab})\e^i-V_\m^{ij}\e_j + \rmi \g\cdot \underline{T} \g_\m
\e^i - \rmi \g_\m
\eta^i  \nn\, ,\\
\d V_\m &=&  -\ft32\rmi \bar\e^{i} \phi_\m^{j} \d_{ij} +4
\bar\e^{i}\g_\m \underline{\chi}^{j}\d_{ij}
  + \rmi \bar\e^{i} \g\cdot \underline{T} \psi_\m^{j} \d_{ij} + \ft32\rmi
\bar\eta^{i}\psi_\m^{j} \d_{ij} \nn\, ,\\
\d V'_\m{}^{ij} &=&  -\ft32\rmi \bar\e^{(i} \phi_\m^{j)} +4
\bar\e^{(i}\g_\m \underline{\chi}^{j)}
  + \rmi \bar\e^{(i} \g\cdot \underline{T} \psi_\m^{j)} + \ft32\rmi 
\bar\eta^{(i}\psi_\m^{j)} +\ft32\rmi \bar\e^{k} \phi_\m^{l} \d_{kl}\d^{ij} \nn \\
&&  -2
\bar\e^{k}\g_\m \underline{\chi}^{l}\d_{kl }\d^{ij}
  -\ft12 \rmi \bar\e^{k} \g\cdot \underline{T} \psi_\m^{l} \d_{kl}\d^{ij} - \ft34\rmi
\bar\eta^{k}\psi_\m^{l} \d_{kl}\d^{ij} \nn\, ,\\
 \d C_\m
&=& -\ft12\rmi \s \bar{\e} \p_\m + \ft12
\bar{\e} \g_\m \p, \nonumber\\
 \d B_{\m\n}
&=& \ft12 \s^2 \bar{\e} \g_{[\m} \p_{\n]} + \ft12 \rmi \s \bar{\e}
\g_{\m\n} \p + C_{[\m} \d(\e) C_{\n]}, \nonumber\\
\d \p^i &=& - \ft14 \g \cdot G \e^i -\ft12\rmi \slashed{\partial} \s
\e^i + \s \g \cdot \underline{T} \e^i -\underline{Y}^{ij}\e_j + \s
\eta^i \,,\nonumber\\
\d \s &=& \ft12 \rmi \bar{\e} \p \, ,\nonumber\\
\delta E_{a} &=& -\tfrac{1}{2\sqrt2} \bar{\e}^i \g_{ab} \g^c  V'_{c (i}{}^k \d_{j)k} \p^{b j} +\ft14 \bar\e \g_{ab} \g^c E_c \p^b  +\ft14\rmi \bar\e \g_{ab} N \p^b   \nn\\
&& - \ft1{2\sqrt2} \rmi \bar\e^i \g_{ab} \g \cdot T \p^{b j} \d_{ij} + \ft3{2\sqrt2}\rmi \bar\e^i \g_{ab} \f^{b j}\d_{ij}\,, \nn\\
\delta N &=& -\tfrac{1}{2\sqrt2} \rmi \bar{\e}^i \g^{a} \g^b  V'_{b (i}{}^k \d_{j)k} \p_a^j +\ft14 \bar\e \g^a \g^b  E_b \p_a -\ft14 \bar\e \g_{a} N \p^a \nn\\
&& + \ft1{2\sqrt2}  \bar\e^i \g^{a} \g \cdot T \p_a^{j} \d_{ij} - \ft3{2\sqrt2} \bar\e^i \g^{a} \f_a^{j}\d_{ij}   - 2\sqrt2  i \bar{\e}^i \underline{\chi}^j \d_{ij} \,, \label{offshelltransfrules}
\eea
where the parameter $\eta^i$ is as described in (\ref{decomS}).

Note that the $U(1)_R$  symmetry of the off-shell supergravity is gauged via the auxiliary $V_\m$ 
\bea
\d_\l V_\m = \partial_\m \l, \quad \d_\l \p_\m^i = \tfrac12 \d^{ij} \l \p_{\m j}, \quad \d_\l \p^i = \tfrac12 \d^{ij} \l \p_{j}\,,
\label{gaugedV}
\eea
where $\l$ is the parameter of the $U(1)_R$ symmetry. Also note that the CS term, which is characteristic of the ${\cal{N}} = 2, D=5$ formulation is hidden inside the term $H_{\m\n\r}H^{\m\n\r}$, and it becomes manifest in the action in the on-shell formalism due to the dualization of $H_{\m\n\r}$ as we shall discuss in the following section. 

Upon going on-shell by solving for the auxiliary fields $E_\m ,\, N,\, V_\m$ and $V_\m^{'ij}$ we obtain ungauged Maxwell-Einstein supergravity as constructed in \cite{Gunaydin:1983bi, Nishino:2000cz}. The elimination of auxiliaries will be discussed in the next section.

\section{Minimal Gauged Supergravity}\label{gaugedtheory}
The off-shell theory discussed in subsection \ref{ss: gaugefix} has two $U(1)$ symmetries: $U(1)_V$ and $U(1)_C$. The first one is related to the auxiliary gauge field $V_{\m}$ and the second one is related to the vector field of the Dilaton Weyl Multiplet $C_{\m}$. The fermions in the theory transform under $U(1)_V$, thus their covariant derivatives contain $V_{\m}$.  This gauging by $V_\m$, however, is undesirable since $V_\m$ has no kinetic term. To obtain a dynamical gauging, one may choose to eliminate $V_\m$ and to promote,  in the resulting on-shell theory, the vector field of the Dilaton Weyl multiplet $C_\m$ to a gauge field by replacing the covariant derivatives of the fermionic fields with $U(1)_C$ covariant derivatives. Such a covariantization in the Lagrangian breaks supersymmetry and hence $g$-dependent gauge invariant terms must be added to the Lagrangian in order to restore supersymmetry \cite{Gunaydin:1984ak}. The necessary $g$-dependent gauge invariant terms can be found by the Noether procedure. However, when considering more complicated scenarios, such as the inclusion of $R^2$ terms, the step-by-step Noether procedure is tedious and cumbersome. Therefore, we devote this section to an easier procedure to construct the internally gauged theory. 

In subsection \ref{ss: offshellgauged} we add to the ungauged off-shell Lagrangian $\mathcal{L}_{0}$ (\ref{VplusL}) a vector-linear coupling  $g\mathcal{L}_{VL}$ (\ref{VLaction1}), where $g$ is a coupling constant. We show that the expressions for the Standard Weyl matter fields, obtained via the field equations of the Vector Multiplet components, get deformed by $g$-dependent terms. In subsection \ref{ss: onshellgauged} we discuss the elimination of the auxiliaries $V_{\mu}, V'{}^{ij}_{\mu}, N, P_{\m}$ and dualize the 2-form $B_{\m\n}$ to a vector $\tilde{C}_{\m}$. We show that the resulting theory describes Einstein-Maxwell supergravity \cite{Gunaydin:1984ak} in which a linear combination of $C_{\m}$ and $\tilde{C}_{\m}$ plays the role of the gauge field. Finally, in subsection \ref{ss: truncation}, we show that we can consistently truncate the matter content that is coupled to supergravity, thereby breaking the gauge group down to a single $U(1)_R$, gauged by $C_{\m}$, and obtain the minimal on-shell gauged supergravity discussed \cite{Gunaydin:1984ak}.

\subsection{The Off-shell Internally Gauged Supergravity} \label{ss: offshellgauged}

Our starting point for the construction of the internally gauged supergravity is the following Lagrangian
\bea
\mathcal{L}_g&\equiv&\mathcal{L}_0-3g\mathcal{L}_{VL} \nn \\
&=&\mathcal{L}_L-3\mathcal{L}_V-3g\mathcal{L}_{VL}\,, \label{coupledL}
\eea
where $\mathcal{L}_L$, $\mathcal{L}_V$ and $\mathcal{L}_{VL}$ are given in (\ref{SCAction}), (\ref{vectoractionf}) and  (\ref{VLaction1}). The field equations for $Y^{ij}, \sigma, \p^i$ and $C_{\mu}$ give rise to the following map between the Standard Weyl Multiplet and the Dilaton Weyl Multiplet 
\begin{eqnarray}
 Y^{ij}_g &=&   \tfrac14  \rmi \s^{-1} \bar\p^i \p^j - \tfrac1{2} g \s^{-1} L^{ij}\,, \label{deformedY} \\ 
\chi^i_g &=& \tfrac18 \rmi \s^{-1}\slashed{\cal{D}}\p^i + \tfrac1{16} \rmi \s^{-2}\slashed{\cal{D}} \s \p^i - \tfrac{1}{32} \s^{-2} \g \cdot G \p^i \nn\\ && + \tfrac14 \s^{-1} \g \cdot T \p^i + \tfrac1{8}\s^{-2}Y^{ij}\p_j+\frac{1}{8}g\sigma^{-2}\vf^i \,, \label{deformedchi} \\
 D_g &=& \tfrac14 \s^{-1} \Box^c \s +\tfrac{1}{8} \s^{-2}( {\cal{D}}_a \s)( {\cal{D}}^a \s) - \tfrac1{16} \s^{-2} G_{\m\n} G^{\m\n} \nn\\
&& - \tfrac{1}{8} \s^{-2} \bar\p{\slashed{\cal{D}}} \p +\tfrac{1}{4}\sigma^{-2}Y^{ij}Y_{ij}- 4\rmi \s^{-1} \bar\p \chi +\tfrac{1}{8}\sigma^{-2}\opsi_{\m}\g_{\n}\p G^{\m\n} \nn\\
&&-\tfrac{1}{16}i\sigma^{-1}\opsi_{\m}\p_{\m}G^{\m\n} + \Big( -\tfrac{26}3 T_{ab} + 2 \s^{-1} \widehat G_{ab} + \tfrac14 \rmi \s^{-2} \bar\p \g_{ab} \p \Big) T^{ab}\nn\\
&& +\tfrac{1}{4}g \sigma^{-2} N\,, \label{deformedD} 
\eea

\bea
{\cal{D}}_{[a}\widehat {\cal{H}}_{bcd]}&=& \tfrac{3}{4}\widehat G_{[ab} \widehat G_{cd]} + \tfrac12 g {\cal{D}}_{[a} E_{bcd]} \,, \label{Bianchi}
\end{eqnarray}
where
\begin{equation}
-\tfrac{1}{6}\varepsilon_{abcde}\widehat{{\cal{H}}}^{edc}=8\sigma^2T_{ab}-\sigma\widehat{G}_{ab}-\tfrac{1}{4}i\opsi\gamma_{ab}\psi\,. \label{deformedH}
\end{equation}
The subscript $g$ in the equations (\ref{deformedY}), (\ref{deformedchi}) and (\ref{deformedD}) indicates that the expressions for $Y^{ij}, \chi^i$ and $D$ now pick up $g$ dependent terms. Note that the expressions for $\chi_g^i$ and $D_g$ are given up to 3- and 4-fermion terms respectively. Comparing the above map with the one in the ungauged case, (\ref{IdenYChiD}) to (\ref{IdenT}), it is clear that the map gets deformed by the gauging. The Bianchi identity (\ref{Bianchi}) implies that
\begin{equation}
\widehat{{\cal{H}}}_{\mu\nu\rho}=3\partial_{[\mu}B_{\nu\rho]}+\tfrac{3}{2}C_{[\mu}G_{\nu\rho]}+ \tfrac12 g E_{\mu\nu\rho} - \tfrac34 \s^2 \bar\p_{[\m} \g_\n \p_{\r]} - \tfrac32 \rmi \s\bar\p_{[\m} \g_{\n\r ]}\p\,. \label{Hfieldstrength}
\end{equation}
The above equation for $\widehat{\cal{H}}_{\m\n\r}$ is clearly not gauge invariant since $E_{\m\n\r}$ has the gauge invariance $\d_{\L} E_{\m\n\r} = 3 \partial_{[\m} \L_{\n\r]}$. In order to balance that out, $B_{\m\n}$ needs to have the additional gauge invariance
\bea
\d_\L B_{\m\n} = 2 \partial_{[\m} \L_{\n ]} - \tfrac12 \L G_{\m\n} - \tfrac12 g \L_{\m\n}\,.
\eea

Using the above expressions for $Y^{ij}$, $D$, $T_{ab}$ and $\chi^i$ in the Lagrangian (\ref{coupledL}) and imposing the gauge fixing conditions ($\ref{gaugefix}$) we obtain the following off-shell Poincar\'e Lagrangian
\begin{eqnarray}
e^{-1}\mathcal{L}_{g, DW}|_{L=-1}&=&\tfrac{1}{2}R-\tfrac{1}{4}\sigma^{-2}G_{\m\n} G^{\m\n}-2g C_{\mu}P^{\mu} -\tfrac{1}{6}\sigma^{-4}{\cal{H}}_{\m\n\r} {\cal{H}}^{\m\n\r}-N^2-gN(\sigma^{-2}+2\sigma) \nn\\
&&-g^2 (\tfrac14 \s^{-4} - \s^{-1} )+P^\m P_\m +\sqrt{2}P^{\mu}V_{\mu}-V'_{\mu}{}^{ij}V'_{\mu ij}-\tfrac{3}{2}\sigma^{-2}\partial_{\mu}\sigma\partial^{\mu}\sigma \nn\\
&&-\tfrac{1}{2} \bar{\p}_{\m} \g^{\m\n\r} D_\n \p_\r -\tfrac{3}{2}\sigma^{-2}\bar{\p}\slashed{D}' \p-\tfrac{3}{2}i\sigma^{-2}\bar{\p}\g^{\m}\g^{\r}\p_{\m}\partial_{\r}\sigma \nn \\
&&-\tfrac{1}{8}i\sigma^{-1}\bar{\p}_{\m}\g^{\m\n\r\s}\p_{\n}G_{\r\s}-\tfrac{1}{4}i\sigma^{-1}\bar{\p}_{\m}\p_{\n}G^{\m\n}-\tfrac{1}{4}\sigma^{-2}\bar{\p}_{\m}\g^{\m\n\r}\p G_{\n\r}  \nn \\
&&+\tfrac{1}{2}\sigma^{-2}\bar{\p}_{\m}\g_{\n}\p G^{\m\n}-\tfrac{1}{8}i\sigma^{-3}\bar{\p}\g\cdot G\p+\tfrac{1}{2\sqrt{2}}\bar\p_\m^i\g^{\m\n\r}\p_\n^j P_\r \d_{ij}\nn \\
&&-\tfrac{1}{24}\sigma^{-2}\bar{\p}_{\m}\g^{\m\n\r\s\l}\p_{\n}{\cal{H}}_{\r\s\l}+\tfrac{1}{4}\sigma^{-2}\bar{\p}_{\m}\g_{\n}\p_{\r}{\cal{H}}^{\m\n\r}\nn \\
&&-\tfrac{1}{6}i\sigma^{-3}\bar{\p}_{\m}\g^{\m\n\r\s}\p {\cal{H}}_{\n\r\s}+\tfrac{1}{2}i\sigma^{-3}\bar{\p}_{\m}\g_{\n\r}\p {\cal{H}}^{\m\n\r}-\tfrac{5}{24}\sigma^{-4}\bar{\p}\g\cdot {\cal{H}} \p \nn\\
&& +\tfrac{1}{4\sqrt{2}}\rmi g  \s^{-2}\bar\p_\m^i \g^{\m\n} \p_\n^j \d_{ij} +\tfrac{1}{2\sqrt{2}}\rmi g  \s \bar\p_\m^i \g^{\m\n} \p_\n^j \d_{ij} - \tfrac{1}{\sqrt2} g \bar\p_\m^i \g^\m \p^j \d_{ij} \nn\\
&&  +\ft1{\sqrt2} g \s^{-3}\bar\p_\m^i \g^\m \p^j \d_{ij}-\tfrac{1}{4\sqrt2}\rmi g \s^{-1} \bar\p^i \p^j \d_{ij}-\tfrac{5}{4\sqrt2}\rmi g \s^{-4} \bar\p^i \p^j \d_{ij}\,, \label{offshellgaugedL}
\end{eqnarray}
where ${\cal{H}}_{\m\n\r}$ is defined as
\bea
{\cal{H}}_{\mu\nu\rho}=B_{\m\n\r}+\tfrac{3}{2}C_{[\mu}G_{\nu\rho]}+ \tfrac12 g E_{\mu\nu\rho}\,. \label{noncovH}
\eea
This Lagrangian is invariant under the transformation rules given in (\ref{offshelltransfrules}), where the underlined fields are now to be evaluated using the deformed expressions, (\ref{deformedY}) to (\ref{deformedH}).

This theory has a $U(1)_V \times U(1)_C$ gauge group parametrized by $\lambda$ and $\eta$ respectively. The fermion covariant derivatives are defined in (\ref{covder}) and contain $V_{\m}$. The gauge transformations of the relevant fields are given by
\bea
\delta_\l V_{\m}&=&\partial_{\m}\lambda\,, \qquad \delta_\eta C_{\m}=\partial_{\m}\eta\,, \nn \\
\delta_\l \psi_{\m}^i&=&\tfrac{1}{2} \lambda\delta^{ij}\psi_{\m j}\,, \qquad \delta_\l \psi^i=\tfrac{1}{2} \lambda\delta^{ij}\psi_{j}\,. \label{gaugetransf}
\eea

\subsection{Einstein-Maxwell Supergravity} \label{ss: onshellgauged}

In this subsection, we will eliminate the auxiliaries $V_{\mu}, V_{\mu}'{}^{ij}, N$ and  $P_{\m}$, present the dualization of the 2-form gauge field $B_{\m\n}$ to a vector field $\tilde{C}_\m$ and discuss the resulting on-shell theory. We will show that the on-shell theory describes Einstein-Maxwell supergravity as constructed in \cite{Gunaydin:1984ak}.

Let us start with the field equations for $N$ and $V'_{\mu}{}^{ij}$
\begin{eqnarray}
0&=&2N+g(\sigma^{-2}+2\sigma)\,,  \\
0&=&V'_{\mu}{}^{ij}+\tfrac34 \s^{-2}\bar\p^i \g_\m \p^j\,.
\end{eqnarray}
Using these two field equations in (\ref{offshellgaugedL}), we obtain the following Lagrangian (up to 4-fermion terms)
\bea
e^{-1}\mathcal{L}_2&=&\tfrac{1}{2}R+g^2(2\sigma^{-1}+\sigma^2)-\tfrac{3}{2}\sigma^{-2}\partial_{\mu}\sigma\partial^{\mu}\sigma-\tfrac{1}{4}\sigma^{-2}G_{\m\n} G^{\m\n} \nn\\
&&-\tfrac{1}{6}\sigma^{-4}{\cal{H}}_{\m\n\r} {\cal{H}}^{\m\n\r}+P^\m P_\m +\sqrt{2}P^{\mu}V_{\mu}-2g C_{\mu}P^{\mu} \nn\\
&&-\tfrac{1}{2} \bar{\p}_{\m} \g^{\m\n\r} D_\n \p_\r -\tfrac{3}{2}\sigma^{-2}\bar{\p}\slashed{D}' \p-\tfrac{3}{2}i\sigma^{-2}\bar{\p}\g^{\m}\g^{\r}\p_{\m}\partial_{\r}\sigma \nn \\
&&-\tfrac{1}{8}i\sigma^{-1}\bar{\p}_{\m}\g^{\m\n\r\s}\p_{\n}G_{\r\s}-\tfrac{1}{4}i\sigma^{-1}\bar{\p}_{\m}\p_{\n}G^{\m\n}-\tfrac{1}{4}\sigma^{-2}\bar{\p}_{\m}\g^{\m\n\r}\p G_{\n\r}  \nn \\
&&+\tfrac{1}{2}\sigma^{-2}\bar{\p}_{\m}\g_{\n}\p G^{\m\n}-\tfrac{1}{8}i\sigma^{-3}\bar{\p}\g\cdot G\p+\tfrac{1}{2\sqrt{2}}\bar\p_\m^i\g^{\m\n\r}\p_\n^j P_\r \d_{ij}\nn \\
&&-\tfrac{1}{24}\sigma^{-2}\bar{\p}_{\m}\g^{\m\n\r\s\l}\p_{\n}{\cal{H}}_{\r\s\l}+\tfrac{1}{4}\sigma^{-2}\bar{\p}_{\m}\g_{\n}\p_{\r}{\cal{H}}^{\m\n\r}\nn \\
&&-\tfrac{1}{6}i\sigma^{-3}\bar{\p}_{\m}\g^{\m\n\r\s}\p {\cal{H}}_{\n\r\s}+\tfrac{1}{2}i\sigma^{-3}\bar{\p}_{\m}\g_{\n\r}\p {\cal{H}}^{\m\n\r}-\tfrac{5}{24}\sigma^{-4}\bar{\p}\g\cdot {\cal{H}} \p \nn\\
&& +\tfrac{1}{4\sqrt{2}}\rmi g  \s^{-2}\bar\p_\m^i \g^{\m\n} \p_\n^j \d_{ij} +\tfrac{1}{2\sqrt{2}}\rmi g  \s \bar\p_\m^i \g^{\m\n} \p_\n^j \d_{ij} - \tfrac{1}{\sqrt2} g \bar\p_\m^i \g^\m \p^j \d_{ij} \nn\\
&&  +\ft1{\sqrt2} g \s^{-3}\bar\p_\m^i \g^\m \p^j \d_{ij}-\tfrac{1}{4\sqrt2}\rmi g \s^{-1} \bar\p^i \p^j \d_{ij}-\tfrac{5}{4\sqrt2}\rmi g \s^{-4} \bar\p^i \p^j \d_{ij}\,, \label{PotLag}
\end{eqnarray}
where
\bea
D_\m \p^i&=&\Bigl(\partial_\m+\tfrac{1}{4}\omega_\m{}^{ab}\g_{ab}\Bigr)\p^i-\tfrac12 V_\m \d^{ij} \p_{j}\,.
\eea

Before proceeding it is useful to dualize the 2-form $B_{\m\n}$ to a vector $\tilde{C}_{\m}$. We do this by adding a Lagrange multiplier term 
\bea
{\cal{L}}'=-\ft{1}{6}\varepsilon^{\m\n\r\s\l} B_{\m\n\r} \partial_\s \tilde{C}_\l\,.
\label{LMultip}
\eea
This Lagrange multiplier introduces another $U(1)$ symmetry in the theory which we will denote $U(1)_{\tilde{C}}$ and which is parametrized by $\tilde{\eta}$. Since the Bianchi identity $\partial^{[\m}B^{\n\r\s ]} = 0$ is now imposed by the field equation for $\tilde{C}_{\m}$, we can treat $B_{\m\n\r}$ as an independent field and compute its field equation. Taking both the Lagrangian (\ref{PotLag}) and the Lagrange multiplier term into account, the field equation for $B_{\m\n\r}$ reads
\bea
{\cal{H}}^{\m\n\r} &=& -\ft12 \s^4e^{-1}\varepsilon^{\m\n\r\s\l} \partial_\s \tilde{C}_\l- \tfrac18 \s^{2}\bar\p_\s \g^{\m\n\r\s\l} \p_\l + \tfrac34\s^{2} \bar\p^{[\m} \g^\n \p^{\r ]} \nn \\
&&+\tfrac{1}{2}i\sigma^{-3}\bar{\p}_{\s}\g^{\m\n\r\s}\p+\tfrac{3}{2}i\sigma\bar{\p}^{\m}\g^{\n\r}\p-\tfrac{5}{8}\bar{\p}\g^{\m\n\r}\p\, \label{EoMB}
\eea
or, using (\ref{noncovH}),
\bea
B^{\m\n\r} &=& {\cal{H}}^{\m\n\r} - \tfrac32 C^{[\m} G^{\n\r ]} -  \tfrac12 g E^{\m\n\r} \nn\\
&=& -\ft12 \s^4e^{-1} \varepsilon^{\m\n\r\s\l} \partial_\s \tilde{C}_{\l}- \tfrac32 C^{[\m} G^{\n\r ]} -  \tfrac12 g E^{\m\n\r}- \tfrac18 \s^{2}\bar\p_\s \g^{\m\n\r\s\l} \p_\l \nn\\
&&+ \tfrac34\s^{2} \bar\p^{[\m} \g^\n \p^{\r ]}+\tfrac{1}{2}i\sigma^{-3}\bar{\p}_{\s}\g^{\m\n\r\s}\p+\tfrac{3}{2}i\sigma\bar{\p}^{\m}\g^{\n\r}\p-\tfrac{5}{8}\bar{\p}\g^{\m\n\r}\p\,. \label{Bmnr}
\eea
From (\ref{EoMB}) we can read of the transformation rule for $\tilde{C}_{\m}$
\bea
\d \tilde{C}_{\m}=-\tfrac{1}{2}i\sigma^{-2}\bar{\e}\p_{\m}+\tfrac{1}{2}\sigma^{-3}\bar{\e}\g_{\m}\p\,.
\eea
Using (\ref{Bmnr}), the Lagrangian now reads (up to 4-fermion terms)
\bea
e^{-1}({\cal{L}}_2+\cal{L}')&=& \tfrac12 R + g^2 (2\s^{-1} + \s^2 ) +P^\m P_\m + \sqrt2 P^\m V_\m - \tfrac32 \s^{-2} \partial_\m \s \partial^\m \s\nn\\
&&-\tfrac14 \s^{-2} G_{\m\n} G^{\m\n} -\ft12  \s^4 \partial_{[\m} \tilde{C}_{\n ]}  \partial^{\m}\tilde{C}^{\n} - 2 g C_\m P^\m + \tfrac14e^{-1} \varepsilon^{\m\n\r\s\l} G_{\m\n} C_{\r} \partial_{\s}\tilde{C}_{\l}\nn\\
 &&+ \tfrac1{12} g e^{-1}\varepsilon^{\m\n\r\s\l} E_{\m\n\r} \partial_{\s}\tilde{C}_{\l} -\tfrac{1}{2} \bar{\p}_{\m} \g^{\m\n\r} D_\n \p_\r -\tfrac{3}{2}\sigma^{-2}\bar{\p}\slashed{D}\p \nn \\
&&-\tfrac{3}{2}i\sigma^{-2}\bar{\p}\g^{\m}\g^{\r}\p_{\m}\partial_{\r}\sigma-\tfrac{1}{8}i\sigma^{-1}\bar{\p}_{\m}\g^{\m\n\r\s}\p_{\n}G_{\r\s}-\tfrac{1}{4}i\sigma^{-1}\bar{\p}_{\m}\p_{\n}G^{\m\n}  \nn \\
&&-\tfrac{1}{4}\sigma^{-2}\bar{\p}_{\m}\g^{\m\n\r}\p G_{\n\r}+\tfrac{1}{2}\sigma^{-2}\bar{\p}_{\m}\g_{\n}\p G^{\m\n}-\tfrac{1}{8}i\sigma^{-3}\bar{\p}\g\cdot G\p\nn \\
&&+\tfrac{1}{2\sqrt{2}}\bar\p_\m^i\g^{\m\n\r}\p_\n^j P_\r \d_{ij}-\tfrac{1}{4}\rmi \sigma^{2}\bar{\p}_{[\m}\p_{\n ]}\partial^\m \tilde{C}^\n -\tfrac{1}{8}\rmi \sigma^{2}\bar{\p}_{\m} \g^{\m\n\r\s}\p_{\n}\partial_\r \tilde{C}_\s \nn \\
&&- \sigma\bar{\p}_{[\m}\g_{\n ]}\p \partial^\m \tilde{C}^\n +\tfrac12 \sigma \bar{\p}_{\m}\g^{\m\n\r}\p \partial_\n \tilde{C}_\r +\tfrac{5}{8}i\bar{\p}\g_{\m\n} \p \partial^\m \tilde{C}^\n\nn \\
&& +\tfrac{1}{4\sqrt{2}}\rmi g  \s^{-2}\bar\p_\m^i \g^{\m\n} \p_\n^j \d_{ij} +\tfrac{1}{2\sqrt{2}}\rmi g  \s \bar\p_\m^i \g^{\m\n} \p_\n^j \d_{ij} - \tfrac{1}{\sqrt2} g \bar\p_\m^i \g^\m \p^j \d_{ij} \nn\\
&&  +\ft1{\sqrt2} g \s^{-3}\bar\p_\m^i \g^\m \p^j \d_{ij}-\tfrac{1}{4\sqrt2}\rmi g \s^{-1} \bar\p^i \p^j \d_{ij}-\tfrac{5}{4\sqrt2}\rmi g \s^{-4} \bar\p^i \p^j \d_{ij}\,.
\label{DualFerm}
\eea

Let us now consider the field equations for $V_\m$ and $E_{\mu\nu\rho}$
\bea
0 &=& P^\m + \tfrac1{4\sqrt2} \bar\p^i_\n \g^{\m\n\r} \p^j_\r \d_{ij}-\tfrac{3}{4\sqrt{2}}\sigma^{-2}\bar{\p}_i\g^{\m}\p_j\d^{ij}\,, \label{EomP} \\
0&=&\varepsilon^{\m\n\r\s\l}\partial_\m\Bigl(P_\n + \tfrac{1}{\sqrt{2}}V_\n-gC_\n-\tfrac{1}{2}g\tilde{C}_\n + \tfrac1{4\sqrt2} \bar\p_i^\t \g_{\n\t\xi} \p_j^\xi \d^{ij} \Bigr)\,. \label{EoMVm}
\eea
The latter equation implies that
\bea
P_{\m}&=&\partial_{\m} \phi-\tfrac{1}{\sqrt{2}}V_{\m}+gC_{\m}+\tfrac{1}{2}g\tilde{C}_{\m}-\tfrac1{4\sqrt2} \bar\p_i^\t \g_{\m\t\xi} \p_j^\xi \d^{ij}, \label{Stueckelberg1}
\eea
where $\phi$ is a Stueckelberg scalar that transforms under $U(1)_V\times U(1)_C\times U(1)_{\tilde{C}}$ as
\bea
\delta_g \phi&=&\tfrac{1}{\sqrt{2}}\lambda-g\eta -\tfrac{1}{2}g\tilde{\eta}\,.
\eea
We can break the gauge group down to $U(1)^2$ by fixing the Stueckelberg scalar to a constant $\phi=\phi_0$ . If we now use (\ref{Stueckelberg1}) in combination with (\ref{EomP}) we obtain
\bea
V_{\m}&=&\sqrt{2}g(C_{\m}+\tfrac{1}{2}\tilde{C}_{\m})-\tfrac{3}{4}\sigma^{-2}\opsi^i\g_{\m}\p^j\d_{ij}\,
\eea
and find a decomposition law for the $U(1)$ parameters
\bea
\lambda=\sqrt{2}g(\eta+\tfrac{1}{2}\tilde{\eta})\,.
\eea 
Using this in the Lagrangian given in (\ref{DualFerm}) we find that the on-shell theory is given, up to 4-fermion terms, by
\bea
e^{-1}{\cal{L}}_{EM}&=& \tfrac12 R + g^2 (2\s^{-1} + \s^2 ) - \tfrac32 \s^{-2} \partial_\m \s \partial^\m \s-\tfrac14 \s^{-2} G_{\m\n} G^{\m\n}  \nn\\
&&-\tfrac{1}{2}\s^4 \partial_{[\m} \tilde{C}_{\n ]}  \partial^{\m}\tilde{C}^{\n}+\tfrac{1}{4}e^{-1} \varepsilon^{\m\n\r\s\l} G_{\m\n} C_{\r} \partial_{\s}\tilde{C}_{\l} \nn \\
&&-\tfrac{1}{2} \bar{\p}_{\m} \g^{\m\n\r} \tilde{\nabla}_\n \p_\r-\tfrac{3}{2}\sigma^{-2}\bar{\p}\tilde{\slashed{\nabla}} \p \nn \\
&&-\tfrac{3}{2}i\sigma^{-2}\bar{\p}\g^{\m}\g^{\r}\p_{\m}\partial_{\r}\sigma-\tfrac{1}{8}i\sigma^{-1}\bar{\p}_{\m}\g^{\m\n\r\s}\p_{\n}G_{\r\s}-\tfrac{1}{4}i\sigma^{-1}\bar{\p}_{\m}\p_{\n}G^{\m\n}  \nn \\
&&-\tfrac{1}{4}\sigma^{-2}\bar{\p}_{\m}\g^{\m\n\r}\p G_{\n\r}+\tfrac{1}{2}\sigma^{-2}\bar{\p}_{\m}\g_{\n}\p G^{\m\n}-\tfrac{1}{8}i\sigma^{-3}\bar{\p}\g\cdot G\p\nn \\
&&-\tfrac{1}{4}\rmi \sigma^{2}\bar{\p}_{[\m}\p_{\n ]}\partial^\m \tilde{C}^\n-\tfrac{1}{8}\rmi \sigma^{2}\bar{\p}_{\m} \g^{\m\n\r\s}\p_{\n}\partial_\r \tilde{C}_\s- \sigma\bar{\p}_{[\m}\g_{\n ]}\p \partial^\m \tilde{C}^\n\nn \\
&& +\tfrac12 \sigma \bar{\p}_{\m}\g^{\m\n\r}\p \partial_\n \tilde{C}_\r +\tfrac{5}{8}i\bar{\p}\g_{\m\n} \p \partial^\m \tilde{C}^\n\nn \\ 
&&+\tfrac{1}{4\sqrt{2}}\rmi g  \s^{-2}\bar\p_\m^i \g^{\m\n} \p_\n^j \d_{ij} +\tfrac{1}{2\sqrt{2}}\rmi g  \s \bar\p_\m^i \g^{\m\n} \p_\n^j \d_{ij}-\tfrac{1}{\sqrt2} g \bar\p_\m^i \g^\m \p^j \d_{ij} \nn\\
&&  +\ft1{\sqrt2} g \s^{-3}\bar\p_\m^i \g^\m \p^j \d_{ij}-\tfrac{1}{4\sqrt2}\rmi g \s^{-1} \bar\p^i \p^j \d_{ij}-\tfrac{5}{4\sqrt2}\rmi g \s^{-4} \bar\p^i \p^j \d_{ij}\,,
\label{Lonshell}
\eea
where
\bea
\tilde{\nabla}_\m \p_\n^i &=& \Big(\partial_\m + \tfrac14 \omega_\m{}^{ab}\g_{ab}\Big)\p_\n^i - \tfrac{1}{\sqrt{2}} g(C_\m+\tfrac{1}{2}\tilde{C}_{\m}) \d^{ij} \p_{\n j}\,, \nn \\
\tilde{\nabla}_\m \p^i &=& \Big(\partial_\m + \tfrac14 \omega_\m{}^{ab}\g_{ab}\Big)\p^i - \tfrac{1}{\sqrt{2}} g(C_\m+\tfrac{1}{2}\tilde{C}_{\m}) \d^{ij} \p_j\,.
\eea
From (\ref{gaugetransf}) we find that $\psi_{\m}^i$ transforms under the remaining gauge symmetry as
\bea
\delta_g \psi_{\m}^i&=&\tfrac{1}{\sqrt{2}}g(\eta+\tfrac{1}{2}\tilde{\eta})\delta^{ij}\p_{\m j}\,, \nn \\
\delta_g \psi^i&=&\tfrac{1}{\sqrt{2}}g(\eta+\tfrac{1}{2}\tilde{\eta})\delta^{ij}\p_j\,. \label{remaininggaugepsi}
\eea

The theory given in (\ref{Lonshell}) describes Einstein-Maxwell Supergravity constructed in \cite{Gunaydin:1984ak}. It consists of the fields $\{e_{\m}{}^a, C_{\m}, \psi_{\m}{}^i, \sigma, \tilde{C}_{\m}, \psi^i\}$ accounting for $20+20$ on-shell degrees of freedom. The supersymmetry transformation rules, up to 3-fermion terms, are
\bea
\d e_\m{}^a   &=&  \ft 12\bar\e \g^a \psi_\m  \nn\, ,\\
\d \psi_\m^i   &=& (\partial_\m+\tfrac{1}{4}\omega_\m{}^{ab}\g_{ab})\e^i-\tfrac{1}{\sqrt{2}}g(C_{\m}+\tfrac{1}{2}\tilde{C}_{\m})\d^{ij}\e_j-\tfrac{1}{6\sqrt{2}}ig(\sigma^{-2}+2\sigma)\g_{\m}\d^{ij}\e_j  \nn\, \\
&&+\tfrac{1}{12}i\sigma^{-1}(\g_{\m}{}^{\n\r}-4\d_{\m}^{\n}\g^{\r})(G_{\n\r}+\sigma^{3}\partial_{[\n}\tilde{C}_{\r]})\e^i\,, \nn \\
\d C_\m&=& -\ft12\rmi \s \bar{\e} \p_\m + \ft12
\bar{\e} \g_\m \p, \nonumber\\
\d \s &=& \ft12 \rmi \bar{\e} \p \,, \nn\\
\d \tilde{C}_{\m}&=&-\tfrac{1}{2}i\sigma^{-2}\bar{\e}\p_{\m}+\tfrac{1}{2}\sigma^{-3}\bar{\e}\g_{\m}\p\,, \nn \\
\d \p^i &=&-\ft12\rmi \slashed{\partial} \s\e^i-\tfrac{1}{12}\g^{\m\n}(G_{\m\n}-2\sigma^{3}\partial_{\m}\tilde{C}_{\n})\e^i+\tfrac{1}{3\sqrt{2}}g(\sigma^2-\sigma^{-1})\delta^{ij}\e_j\,. \label{EMtransf}
\eea
The theory has a $U(1)\times U(1)$ gauge symmetry parametrized by $\eta$ and $\tilde{\eta}$. The gauge transformation rules for the gauge vectors and the fermions are given in (\ref{gaugetransf}) and (\ref{remaininggaugepsi}) respectively.

\subsection{Truncation to Minimal Gauged On-Shell Supergravity} \label{ss: truncation}
In this subsection we show that we can consistently truncate the fields $(\s, \p^i, \tilde{C}_\m )$ to obtain on-shell pure gauged $D=5$, ${\cal{N}}=2$ supergravity \cite{Gunaydin:1984ak}. 

Consider the field equation for $\sigma$
\begin{eqnarray}
0&=&3\sigma^{-2}\Box\sigma-3\sigma^{-3}\partial_{\mu}\sigma\partial^{\mu}\sigma+\tfrac{1}{2}\sigma^{-3}G_{\m\n} G^{\m\n}-2\s^3 \partial_{[\m}\tilde{C}_{\n ]}\partial^{\m}\tilde{C}^{\n}\nn\\
&&+\tfrac18 \rmi \s^{-2} \bar\p_\m \g^{\m\n\r\s} \p_\n G_{\r\s} + \tfrac14 \rmi \s^{-2} \bar\p_\m \p_\n G^{\m\n} -\ft12\rmi \s \bar\p_{[\m} \p_{\n ]}\partial^\m \tilde{C}^\n \nn\\
&&- \tfrac14 \rmi \s \bar\p_\m \g^{\m\n\r\s} \p_\n \partial_{\r} \tilde{C}_\s+2g^2(\s-\s^{-2})-\tfrac{1}{2\sqrt{2}}\rmi g  \s^{-3}\bar\p_\m^i \g^{\m\n} \p_\n^j \d_{ij} \nn\\
&&+ \tfrac{1}{2\sqrt{2}}\rmi g \bar\p_\m^i \g^{\m\n} \p_\n^j \d_{ij}+(\psi^i\text{-terms})\, \label{EOMs} 
\end{eqnarray}
and the field equation for $\p^i$
\bea
0 &=& -\ft32 \rmi \s^{-2} \g^\m \g^\n \p_\m^i \partial_\n \s -\ft14 \s^{-2} \g^{\m\n\r} \p_\m^i G_{\n\r} + \ft12 \s^{-2} \g_\m \p_\n^i G^{\m\n} \nn\\
&& +\ft12 \s \g^{\m\n\r} \p_\m^i  \partial_\n \tilde{C}_\r - \s\g_{[\m} \p_{\n ]}^i \partial^\m \tilde{C}^\n  + \tfrac{1}{\sqrt2} g  \g_\m \p_j^\m \d^{ij}\nn\\
&& -\ft1{\sqrt2} g \s^{-3} \g_\m \p_j^\m \d^{ij}+(\psi^i\text{-terms})\,,
\label{EOMp}
\eea
both up to 4-fermion terms. From these equations and from the transformation rules of $\sigma$ and $\psi^i$ in (\ref{EMtransf}), we observe that one can consistently eliminate the matter fields $(\s, \p^i, \tilde{C}_\m )$ by setting $\s=1$, $\p^i=0$ and $\tilde{C}_\m-C_\m=\partial_{\m}a$, where $a$ is a Stueckelberg scalar. The gauge transformation of $a$ is given by
\bea
\d_g a=\eta-\tilde{\eta}\,.
\eea
We can break the $U(1)\times U(1)$ gauge symmetry down to $U(1)$ by setting $a$ to a constant $a=a_0$. This implies
\bea
C_{\m}=\tilde{C}_{\m}\,, \qquad \eta=\tilde{\eta}\,.
\eea

Performing this truncation in (\ref{Lonshell}) we obtain the on-shell Lagrangian for pure gauged $D=5$, ${\cal{N}}=2$ supergravity
\bea
e^{-1}{\cal{L}}_{EM}|_{\sigma=\text{1}, \psi=\text{0}}&=& \tfrac12 R + 3g^2  -\tfrac38  G_{\m\n} G^{\m\n}+ \tfrac18 e^{-1}\varepsilon^{\m\n\r\s\l} C_{\m}G_{\n\r}G_{\s\l}    \nn\\
&&-\tfrac{1}{2} \bar{\p}_{\m} \g^{\m\n\r} \nabla_\n \p_\r -\tfrac38 \rmi \bar\p_\m \p_\n G^{\m\n}-\tfrac3{16} \rmi \bar\p_\m \g^{\m\n\r\s}\p_\n G_{\r\s} \nn\\
&& + \tfrac{3}{4\sqrt{2}}\rmi g\bar\p_\m^i \g^{\m\n} \p_\n^j \d_{ij}\,,
\label{PreFull2}
\eea
where we defined
\bea
 \nabla_\m \p_\n^i = \Big(\partial_\m + \tfrac14 \omega_\m{}^{ab}\g_{ab}\Big)\p_\n^i - \tfrac{3}{2\sqrt{2}} g C_\m \d^{ij} \p_{\n j}\,.
\label{Nabla}
\eea
This Lagrangian is invariant under the transformation rules for $e_{\m}^a$, $\psi_{\m}^i$ and $C_{\m}$ given in (\ref{EMtransf}) with $\sigma=1$, $\p^i=0$ and $\tilde{C}_{\m}=C_{\m}$.

The Lagrangian (\ref{PreFull2}) agrees completely with the result obtained in \cite{Gunaydin:1984ak}. Another off-shell action is obtained in \cite{Bergshoeff:2004kh, Hanaki:2006pj} by coupling the Standard Weyl Multiplet to a Hypermultiplet compensator. After elimination of the auxiliaries and decoupling of the matter fields, these results also agree with ours.


\section{Conclusions}
In this paper we have constructed the complete (bosonic and fermionic) off-shell action for minimal gauged $D=5$, ${\cal{N}}=2$ Poincar\'e supergravity. We obtained this action by using the methods of superconformal tensor calculus. We first constructed the pure (ungauged) theory by constructing the superconformal action for a Linear Multiplet in the background of the Standard Weyl Multiplet (\ref{SCAction}). This action is trivial unless we add the superconformal action of a Vector Multiplet, also in the Standard Weyl background. Eliminating the Standard Weyl matter fields via the field equations of the Vector Multiplet components, we arrived at the superconformal action for the Linear Multiplet in a Dilaton Weyl background. By fixing the redundant symmetries ($D$, $K$, $S$) we obtained the pure off-shell Poincar\'e action (\ref{ActionBos}). 

In a next step we used the internal vector of the Dilaton Weyl Multiplet to gauge the theory. We started again with the sum of a Linear and Vector Multiplet action, both in a Standard Weyl background, and added a superconformal coupling proportional to $g$ between the Linear and Vector Multiplets. Eliminating again the Standard Weyl matter fields via the field equations of the Vector Multiplet components, we observed that the original expressions get deformed by $g$-dependent terms. After fixing the redundant symmetries we obtained the off-shell action for minimal gauged $D=5$, ${\cal{N}}=2$ Poincar\'e supergravity (\ref{offshellgaugedL}). After solving for the auxiliaries and dualizing the 2-form $B_{\m\n}$ to a vector $\tilde{C}_{\m}$, we showed that the resulting theory (\ref{Lonshell}) is Einstein-Maxwell supergravity as constructed in \cite{Gunaydin:1984ak} via the Noether method. We also showed that the field equations and transformation rules allow to truncate $\sigma$, $\psi^i$ and $\tilde{C}_{\m}$. After truncation we ended up with the action of pure on-shell supergravity \cite{Gunaydin:1984ak}.

It would be interesting to investigate the extension of our result with the Weyl squared off-shell invariant $\mathcal{L}_{W^2}$ constructed in \cite{Hanaki:2006pj}. This superconformal invariant is constructed in the Standard Weyl background and can be added to our result to yield
\begin{equation}
\mathcal{L}_L+\mathcal{L}_V+g\mathcal{L}_{VL}+\tfrac{1}{M^2}\mathcal{L}_{W^2},
\end{equation}
with $1/M^2$ some arbitrary parameter. Computing the field equations for the vector field components we obtain expressions for the Standard Weyl fields that now also get contributions proportional to $1/M^2$. Using these expressions and gauge fixing the redundant symmetries we obtain a higher derivative extension of our gauged result. It is interesting to note that the CS term $C \wedge \text{tr}(R \wedge R)$ now contains the internal vector of the Dilaton Weyl Multiplet. Since this resulting theory is in the Dilaton Weyl background it is interesting to study if it is possible to further extend this theory by including the Riemann squared invariant constructed in \cite{Bergshoeff:2011xn}.

Another interesting problem is to study the matter couplings of our new off-shell Poincar\'e theory. It is well known that the $D=5$, ${\cal{N}}=2$ Poincar\'e supergravity coupled to $n$ Vector Multiplets goes under a specific geometrical structure, the so called `very special geometry' \cite{Gunaydin:1983bi, deWit:1992cr}. In rigid theory, the scalars of the Vector Multiplets stay unconstrained. In supergravity, however, the gauge fixing condition for the dilatation invariance restricts the scalars to form a very special real manifold. Similarly, when considering the Hypermultiplet couplings of the theory, the scalars parametrize a hyper-K\"{a}hler manifold for rigid supersymmetry. In supergravity, the gauge fixing conditions remove four compensating scalars resulting in a quaternionic K\"{a}hler geometry \cite{Bergshoeff:2004kh, Bergshoeff:2002qk}.  In our construction, however, the gauge fixing takes place via either the elements of the Dilaton Weyl Multiplet or the elements of the Linear Multiplet. Hence, it would be interesting to study how the constraint on the scalars of the Vector Multiplets or Hypermultiplet arises and to show how our off-shell formulation gives rise to the same on-shell theory as constructed in \cite{ Ceresole:2000jd, Bergshoeff:2004kh}.

\section*{Acknowledgments}
We are grateful to Toine van Proeyen and Ergin Sezgin for carefully reading the draft version of the paper and making many helpful suggestions. M.O. would like to thank the Institute for Theoretical Physics of KU Leuven and the Physics Department of Koc University for hospitality while the paper was written. 
This work is supported in part by the FWO - Vlaanderen, Project No.
G.0651.11,  in part by the Federal Office for Scientific, Technical
and Cultural Affairs through the ``Interuniversity Attraction Poles
Programme -- Belgian Science Policy'' P7/37 P7/37 and in 
part by NSF grant PHY-0906222.

\end{document}